\documentclass[preprint,12pt,authoryear]{elsarticle}

\usepackage{amssymb}
\usepackage{array}
\usepackage{rotating}
\usepackage{xcolor}

\journal{Computerized Medical Imaging and Graphics}

\begin{document}

\begin{frontmatter}



\title{A Survey on Automated Diagnosis of Alzheimer's Disease Using Optical Coherence Tomography and Angiography}


\author[Isik University]{Yasemin Turkan}

\affiliation[Isik University]{organization={Isik University, Department of Computer Engineering},
            addressline={Mesrutiyet Koyu, Universite Sok. No:2, Sile}, 
            city={Istanbul},
            postcode={34980}, 
            country={Turkey}}

\author[Istanbul Technical University]{F. Boray Tek}

\affiliation[Istanbul Technical University]{organization={Istanbul Technical University, Department of Artificial Intelligence and Data Engineering},
            addressline={ITU Ayazaga Kampusu, Maslak}, 
            city={Istanbul},
            postcode={34469}, 
            country={Turkey}}

\begin{abstract}
Retinal optical coherence tomography (OCT) and optical coherence tomography angiography (OCTA) are promising tools for the (early) diagnosis of Alzheimer's disease (AD). These non-invasive imaging techniques are cost-effective and more accessible than alternative neuroimaging tools. However, interpreting and classifying multi-slice scans produced by OCT devices is time-consuming and challenging even for trained practitioners. 

\noindent There are surveys on machine learning and deep learning approaches concerning the automated analysis of OCT scans for various diseases such as glaucoma. However, the current literature lacks an extensive survey on the diagnosis of Alzheimer's disease or cognitive impairment using OCT or OCTA. This has motivated us to do a comprehensive survey aimed at machine/deep learning scientists or practitioners who require an introduction to the problem. The paper contains 1) an introduction to the medical background of Alzheimer's Disease and Cognitive Impairment and their diagnosis using OCT and OCTA imaging modalities, 2) a review of various technical proposals for the problem and the sub-problems from an automated analysis perspective, 3) a systematic review of the recent deep learning studies and available OCT/OCTA datasets directly aimed at the diagnosis of Alzheimer's Disease and Cognitive Impairment. For the latter, we used Publish or Perish Software to search for the relevant studies from various sources such as Scopus, PubMed, and Web of Science. We followed the PRISMA approach to screen an initial pool of 3073 references and determined ten relevant studies (N=10, out of 3073) that directly targeted AD diagnosis. We identified the lack of open OCT/OCTA datasets (about Alzheimer's disease) as the main issue that is impeding the progress in the field.
\end{abstract}

\begin{graphicalabstract}
\includegraphics{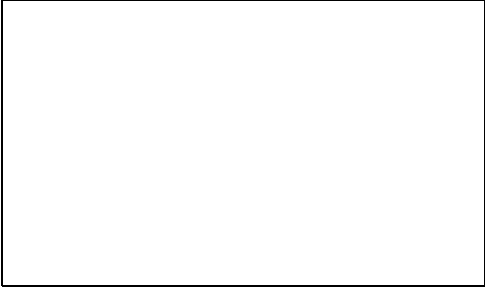}
\end{graphicalabstract}

\begin{highlights}
\item A comprehensive survey on machine and deep learning-based approaches to the diagnosis of Alzheimers's disease in OCT/OCTA scans.
\item A review of the research papers between 2015-2022.
\item A review of the existing and ongoing OCT/OCTA dataset collection initiatives
\end{highlights}

\begin{keyword}
Alzheimer's disease \sep cognitive impairment \sep datasets \sep  dementia \sep deep learning \sep machine learning  \sep OCT \sep OCTA \sep optical coherence tomography \sep optical coherence tomography and angiography \sep retinal imaging 

\end{keyword}

\end{frontmatter}


\section{Introduction}

Alzheimer's disease (AD) is the most common form of dementia. It is an irreversible, progressive brain disorder usually marked by a decline in cognitive functioning with no known treatment. It is characterized by a massive decrease in brain size (neurodegeneration) caused by the neurons' accumulation of proteins (amyloid-beta and tau). Since the retina and the brain grow from the same neural tube, eyes are considered extensions of the brain. Postmortem studies in AD also highlighted the accumulation of amyloid-beta and tau proteins in the retina. More recently, high-resolution visual imaging techniques, including optical coherence tomography (OCT) and optical coherence tomography angiography (OCTA), have been proposed to evaluate structural and vascular changes in the retina of AD patients. 

OCT and OCTA generate large-scale, high-dimensional and multimodal retinal scans, which require tedious and time-consuming readings to be performed by experts. As in similar medical diagnostic fields, machine or deep learning-based automated approaches can significantly enhance physicians' ability to efficiently diagnose abnormalities in the retina.


There are surveys studying the deep learning-based analysis of eye diseases such as glaucoma \citep{Mirzania2021}. 
\cite{Bourkhime2022} recently reviewed machine learning (ML) and novel ophthalmologic biomarkers for AD screening in a brief study. They have identified 13 studies. However, the literature lacks a broader survey on using deep learning (DL) for Alzheimer's disease diagnosis from OCT and OCTA. This has motivated the current comprehensive survey aimed at machine/deep learning scientists or practitioners who require an introduction to the problem.

In this work, we present 1) an introduction to the medical background of Alzheimer's Disease and Cognitive Impairment and their traces that can be observed through OCT and OCTA imaging, 2) a review of various technical proposals for the problem and the sub-problems from an automated analysis perspective, 3) a systematic review of the recent deep learning studies and available OCT/OCTA datasets directly aimed at the diagnosis of Alzheimer's Disease and Cognitive Impairment.
We discuss the progress in the field and future directions in light of existing studies and recent data collection initiatives. 

\begin{figure*}
\centering
\includegraphics[width=1.0\textwidth]{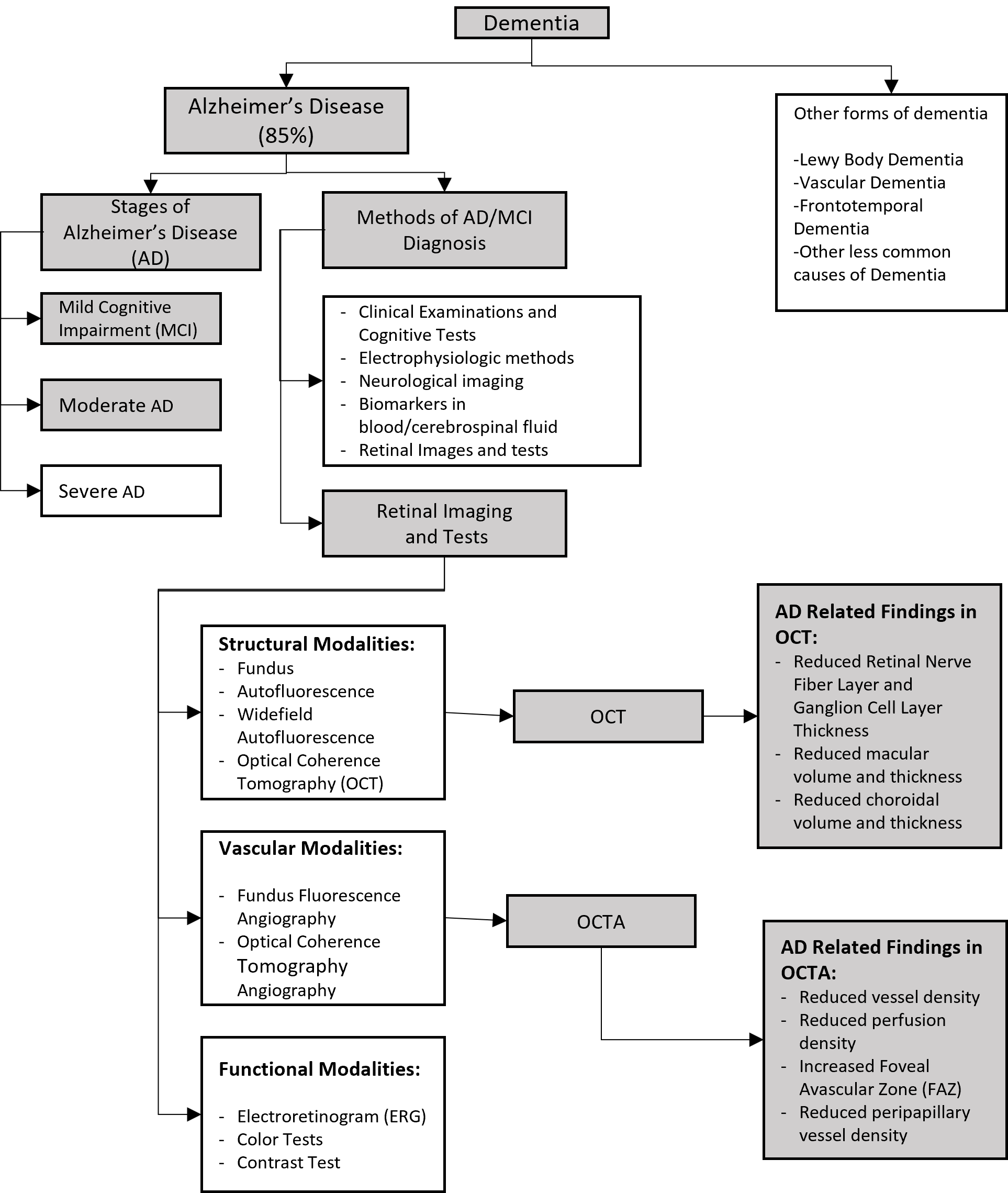}
\caption{Main flow of the medical background and the focus of the study (grey, filled). Common medical findings related to AD are listed under the OCT/OCTA boxes.}
\label{fig:medical_flow}       
\end{figure*}

One of the main goals of this study is to compile a useful survey that can serve as an introduction to the problem. The organization of the paper is geared towards this goal. 
The related work is divided into two parts to explain the medical and technical background of the problem. Section \ref{sec:Medical_Background} explains various biomarkers that were identified by medical researchers to screen, diagnose, or classify AD from OCT and OCTA scans. Figure \ref{fig:medical_flow} summarizes the medical background and highlights our survey's focus of interest. The machine or deep learning approaches or applications to analyze OCT/OCTA scans are examined in Section \ref{sec:Technical}. The guidelines of the Preferred Reporting Items for Systemic Review and Meta-Analysis (PRISMA) \citep{Page2021} are followed in this study, as explained in Section \ref{sec:methods}. The results of the systematic search can be found in Section \ref{sec:Results}. Collecting sufficient, high-quality and uniformly annotated data to build high-accuracy models is difficult. Section \ref{sec:dataset} explains the issues about collecting AD-related OCT data and currently available datasets for the research. Finally, our discussion and conclusions can be found in Sections \ref{sec:discussion} and \ref{sec:conclusion}.

\section{Medical Background}
\label{sec:Medical_Background}
World Health Organization (WHO) classifies dementia as the seventh leading cause of death among older people globally and recognizes it as a public health priority \citep{WHO2019}. WHO defines dementia as follows: 

\textit{Dementia is a syndrome due to disease of the brain, usually of a chronic or progressive nature, in which there is disturbance of multiple higher cortical functions, including memory, thinking, orientation, comprehension, calculation, learning capacity, language, and judgment}
\citep{WorldHealthOrganization}

Dementia is not a single disease \citep{Dening2015}; it is a general term --like heart disease-- that covers a wide range of specific medical conditions (see Figure \ref{fig:medical_flow}). 

\subsection{Alzheimer's Disease (AD)}
\begin{figure}
\centering
\includegraphics[width=0.7\textwidth]{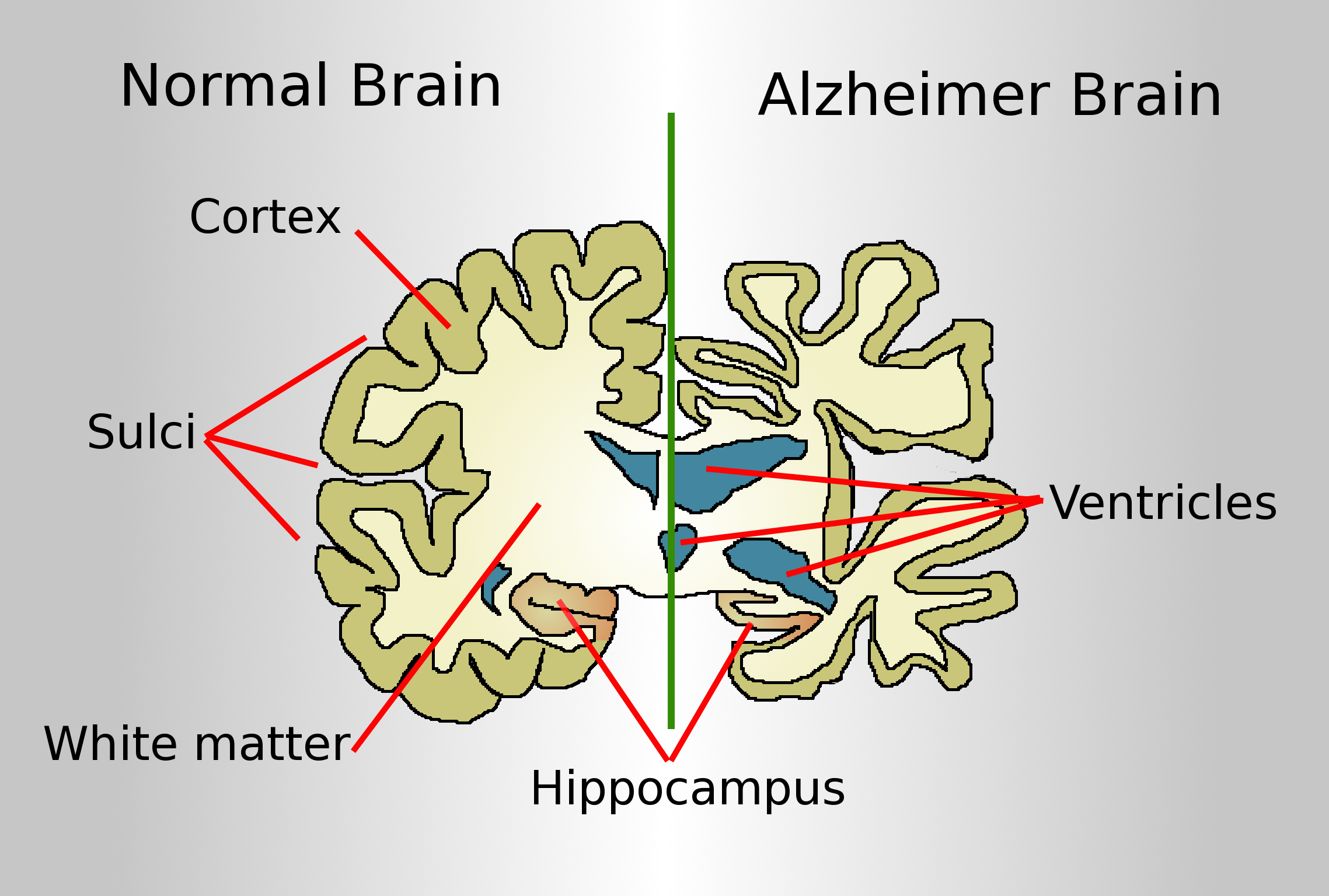}
\caption{AD-affected areas of the brain \citep{Garrondo}. The cerebral cortex and the hippocampus shrink, the ventricles enlarge with cerebral fluid.}
\label{fig:ad-brain2}       
\end{figure}

Alzheimer's disease (AD) is the most common form of dementia and one of the top ten leading causes of death in the world. The slow progress of the disease begins almost 20 years before the symptoms arise \citep{Gaugler2022}. Accumulation of amyloid-beta and tau proteins in the neurons \citep{Dening2015} causes degenerative changes and mass decrease in the brain, especially in the hippocampus, cerebral cortex, and ventricles \citep{HabibHavoutis2017,Mu1999,Juottonen1999} (Figure \ref{fig:ad-brain2}).

AD first starts in the hippocampus region of the brain, which is responsible for forming new memories. The patients begin to have difficulties creating new memories \citep{Gaugler2022}. They may function independently even though the symptoms can be noticed by their families, close friends, and doctors. This phase is generally called Mild Cognitive Impairment (MCI), representing the preclinical stage of AD. Although the clinical diagnosis of MCI is not a necessary precursor to AD, it serves as a risk factor for AD progression \citep{HabibHavoutis2017}. Therefore, differentiating stable MCI (sMCI) from progressive, prodromal MCI (pMCI) is important, and it has been the focus of various studies \citep{Ellendt2017,Yamanakkanavar2020}. 

After the hippocampus, the cerebral cortex responsible for language and information processing begins to thin as the brain gradually shrinks. The patient starts to have difficulty in thinking and planning. Older memories are lost. As more and more neurons die, ventricles filled with cerebrospinal fluid grow larger. This stage is called Alzheimer's disease (Moderate), typically the longest stage that can last many years. As the disease progresses, the person with AD will require greater care \citep{Gaugler2022}. 

In the final stage of the disease (Severe), dementia symptoms are severe. As memory and cognitive skills continue to worsen, significant personality changes may occur, and individuals need extensive care \citep{Gaugler2022}.

\subsection{Diagnosis of AD and MCI}
\label{AD_Diagnosis}

Medical history is the first step and very critical for any clinical diagnosis. Therefore, a detailed history such as family medical history, existing or past medical issues, current health parameters, and cognitive status must be obtained from the patient or the caregiver. This step is vital for understanding the progress of the disease as well. Ruling out other sources of dementia is equally important \citep{HabibHavoutis2017,McKhann2011}. Mini-Mental State Exam (MMSE) is a small and simple test that can easily be applied during clinical examinations. It is the most widely used and globally accepted cognitive test \citep{Folstein1975}. It has five parts: orientation, registration, attention, recall, and language. Other tests such as Alzheimer's Disease Assessment Scale – Cognitive sub-scale (ADAS–Cog), Wechsler Memory Scale, the General Practitioner Assessment of Cognition (GPCOG), Mini-Cog, the Memory Impairment Screen (MIS), the Clock Drawing Test (CDT), and the Montreal Cognitive Assessment (MoCA) \citep{HabibHavoutis2017} can be used independently or in combination with MMSE.

If any form of dementia is suspected, further diagnostic tests follow the clinical examination. Electroencephalography (EEG) and Magnetoencephalography (MEG) are two non-invasive methods used in diagnosing cognitive impairment. EEG measures the brain's electrical activity using electrodes or sensors placed on the scalp. In contrast, MEG measures the magnetic fields generated by the electrical activity using superconductive quantum interference devices placed over the head \citep{Budson2011}. EEG and MEG signals can detect neurodegeneration-induced alterations in synaptic and neural activity. Therefore, these tools provide valuable information about the changes in brain dynamics due to AD \citep{Al-Nuaimi2021}.

Standard skull X-ray examination is ineffective in displaying AD-related changes in soft brain tissues. Computed tomography (CT) and ordinary magnetic resonance imaging (MRI) are more capable than X-ray, albeit with various limitations in diagnosing AD. For example, although CT effectively detects advanced stages of AD, it usually shows no abnormal findings in the early stages. Standard head MRI displays soft tissues (grey matter and white matter) in detail. The brain's size and volume can easily be used to differentiate AD patients from health cohorts (HC). However, MRI does not reveal many AD-related physiological and fluid biomarker changes. Therefore, the specificity and sensitivity of MCI vs. HC diagnosis using head MRI are not as high as needed \citep{Zeng2021, Yamanakkanavar2020}. Modern neuroimaging techniques such as positron emission computed tomography (PET) and functional MRI (fMRI) are more accurate but more expensive alternatives for early AD (MCI) diagnosis. PET is an invasive nuclear medicine imaging technique that helps visualize the cerebral blood flow and functional metabolism of the brain tissue. fMRI examines brain activity and neural network function by detecting changes caused by blood flow. fMRI is mainly used in clinical trials of AD \citep{Zeng2021}.

A$\beta$ plaques, tau proteins, senile plaques (SP), and neurofibrillary tangles (NFTs) are the causes of cell death and dementia according to the amyloid cascade hypothesis \citep{Armstrong2011}. Therefore, these are the most common biomarkers used to diagnose AD in clinical trials. These biomarkers can be found in the cerebrospinal fluid (CSF) or blood. Results obtained from biomarker-rich CSF give more accurate values than the peripheral blood \citep{Nguyen2020}. 

Electrophysiologic methods, neurological imaging, and biomarkers support the final decision. Recently, there has been increased interest in retinal examination as a promising candidate for early AD diagnosis.

\begin{figure}
\centering
\includegraphics[width=0.85\textwidth]{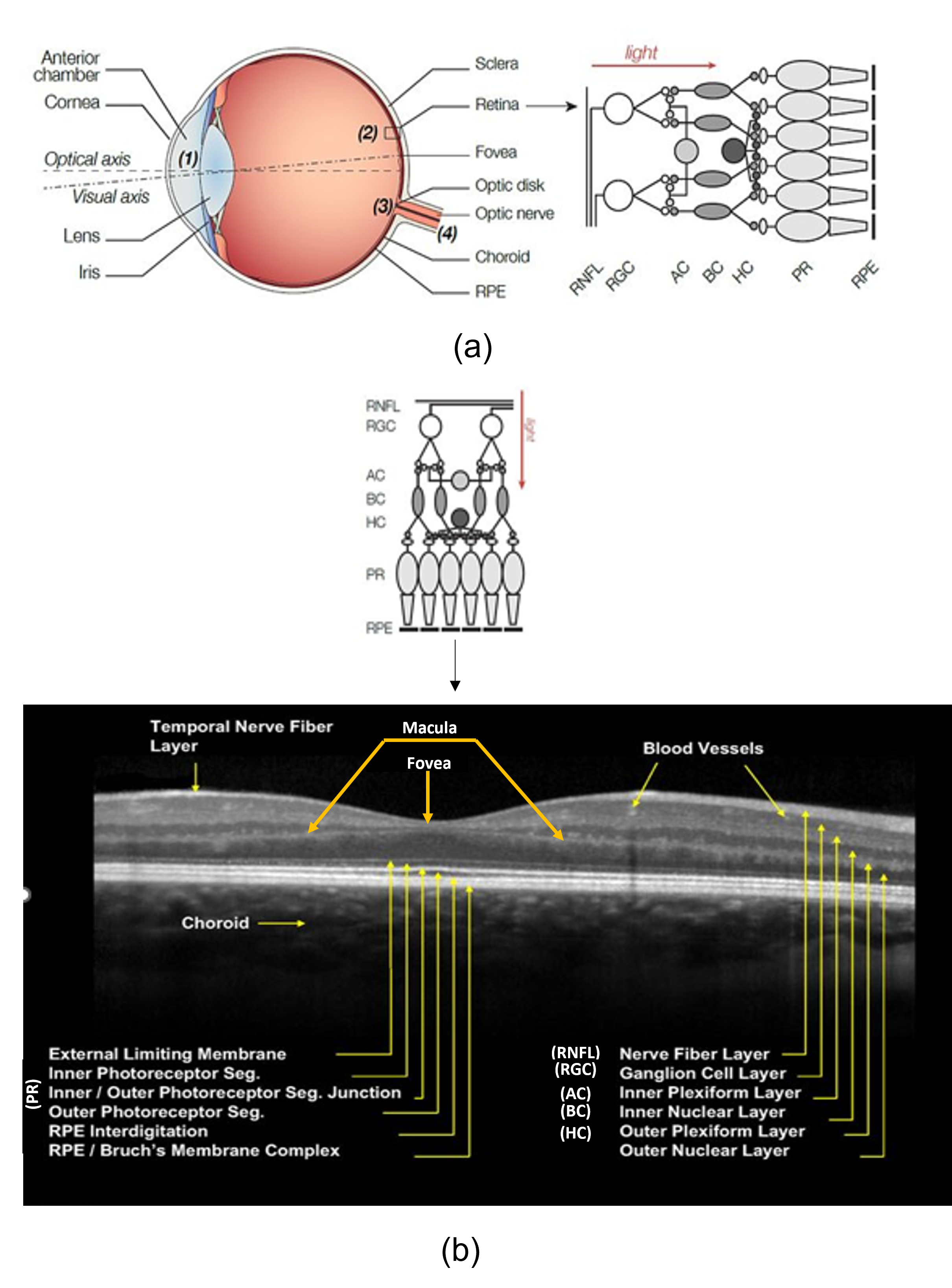}
\caption{(a) Diagram of the eye, the retina, and location of various retinal implants. Retinal layers from bottom to top: retinal pigment epithelium (RPE), photoreceptors (PR), horizontal cells (HC), bipolar cells (BC), amacrine cells (AC), ganglion cells (RGC), nerve fiber layer (RNFL) \citep{Palanker}\\
(b) Raw image taken from OCT and marked by \cite{Heidelberg}. The different layers are marked on the right. Nerve Fiber, Ganglion Cell, and Inner Plexiform Layers exhibit changes in neurodegenerative diseases such as AD.}
\label{fig:retina_oct}       
\end{figure}

During embryonic development, the retina and the optic nerve grow from the neural tube and are thus considered as parts or extensions of the central nervous system \citep{Blazes}. Figure \ref{fig:retina_oct} shows the neural layers of the retina. The only light-sensitive cells in the retina are the photoreceptors. When the light falls on the photoreceptors, the signal is carried from bipolar cells to ganglion cells that fire action potentials in response to the light. All other cells are influenced by light via direct and indirect synaptic interactions. These impulses propagate down the optic nerve to the rest of the brain. The only output source from the retina is the ganglion cells \citep{Bear2016}. 

The eye is the only organ in which we can see the cardiovascular blood vessels and an inner blood-retinal barrier (a tight protective layer of cells and capillaries that prevent larger molecules from entering the retina) similar to the blood-brain barrier \citep{London2013}. Consequently, retinal images are also used to analyze systemic diseases such as cardiovascular disease \citep{Wagner2020}. Since the eye is more accessible than the brain, techniques used to examine the eye are more practical, generally non- or less invasive and more cost-effective than the brain imaging alternatives (MRI, PET, CT) \citep{Song2021, Cunha2022}. The neurodegenerative process that impacts the brain neurons also affects the retinal neurons. Specifically, postmortem AD studies highlighted the accumulation of A$\beta$ and NFTs in the retinal and optic nerve tissue. Moreover, animal \citep{Hadoux2019, Carelli2017, Gardner2020} and human studies \citep{DenHaan2018} conclusively reveal the retinal A$\beta$ accumulations. This raises the question: Can ophthalmological assessments detect Alzheimer's disease and related disorders?

Ophthalmologists commonly use multiple modalities for the assessment of neurodegenerative diseases. These can be grouped into three main categories \citep{Attiku2021}:

\begin{enumerate}
    \item Structural modalities: Fundus, Optical Coherence Tomography (OCT), Autofluorescence \citep{Attiku2021},  Widefield Autofluorescence \citep{Alber2020}.
    \item Vascular modalities: Fundus fluorescence angiography and Optical Coherence Tomography Angiography (OCTA) \citep{Attiku2021}.
    \item Functional modalities: electroretinogram (ERG) \citep{Tsang2018}, various color and contrast tests \citep{London2013}.
    
\end{enumerate}

Recently \cite{Vij2022} have done a systematic survey of advances in retinal modalities for the diagnosis of Alzheimer's disease. They have reviewed structural imaging modalities, functional modalities, and multimodal/paired imaging. Among all of these modalities, OCT is the modality most commonly used (with a rate of 57\%) in AD diagnosis since it is non-invasive, cost-efficient and easy to use. OCTA is a relatively new modality that is receiving increasing attention. Various studies combine OCT and OCTA to improve the performance of AD diagnosis \citep{Wisely2022, Salobrar-Garcia2019, Querques2019}. OCT and OCTA have also been used in diagnosing various cognitive impairments \citep{Chalkias2021a,Ge2021,Lee2018,Hui2021,Fereshetian2021,MacGillivray2018,Romaus-Sanjurjo2022, Yeh2022,Kim2022}. The focus of this survey, OCT/OCTA, are the most promising retinal imaging techniques that provide practical and cost-effective diagnostic for AD and cognitive impairment. However, neither is mature enough yet. 

\subsubsection{Optical coherence tomography (OCT)}
\label{OCT}

OCT is a non-invasive method used to obtain views of retinal structures in two-dimensional (2D), cross-sectional and three-dimensional (3D) volumetric images \citep{Snyder2020}. OCT provides extensive information regarding retinal morphology and assists in diagnosing many diseases (Figure \ref{fig:retina_oct}). It generates structural images by detecting interference signals from the reference mirror and the backscattering signals from the biological tissue sample (see Figure \ref{fig:full_OCT}). It performs micrometer-resolution depth and density measurements. Time domain (TD-OCT), spectral domain (SD-OCT) and swept source (SS-OCT) are the three main types of OCT. They differ in their scanning speeds: in terms of A-scans per second (TD-OCT : 400, SD-OCT: 20,000-40,000, SS-OCT : 100,000-400,000) \citep{Song2021}.

The last three-four years has witnessed an explosion in AD-related research studies using OCT. \cite{Song2021} have done an extensive literature review on diagnosing Alzheimer's disease using OCT imaging modalities. A common theme in these studies is to use quantitative parameters derived from OCT scans of the retina instead of using 3D scans directly. The following are the most common findings we compiled from OCT-based AD diagnosis literature:

\begin{figure}
\centering
\includegraphics[width=0.7\textwidth]{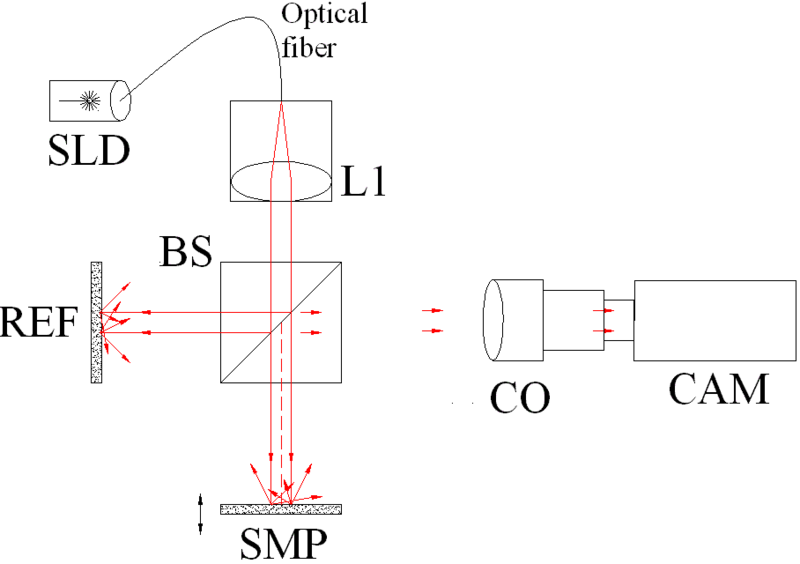}
\caption{Full-field OCT optical setup. Components include super-luminescent diode (SLD), convex lens (L1), 50/50 beamsplitter (BS), camera objective (CO), CMOS-DSP camera (CAM), reference (REF), and sample (SMP). The camera functions as a two-dimensional detector array, and with the OCT technique facilitating scanning in depth, a non-invasive three-dimensional imaging device is achieved \citep{Pumpkinegam}.}
\label{fig:full_OCT}       
\end{figure}

\begin{figure}
\centering
\includegraphics[width=0.96\textwidth]{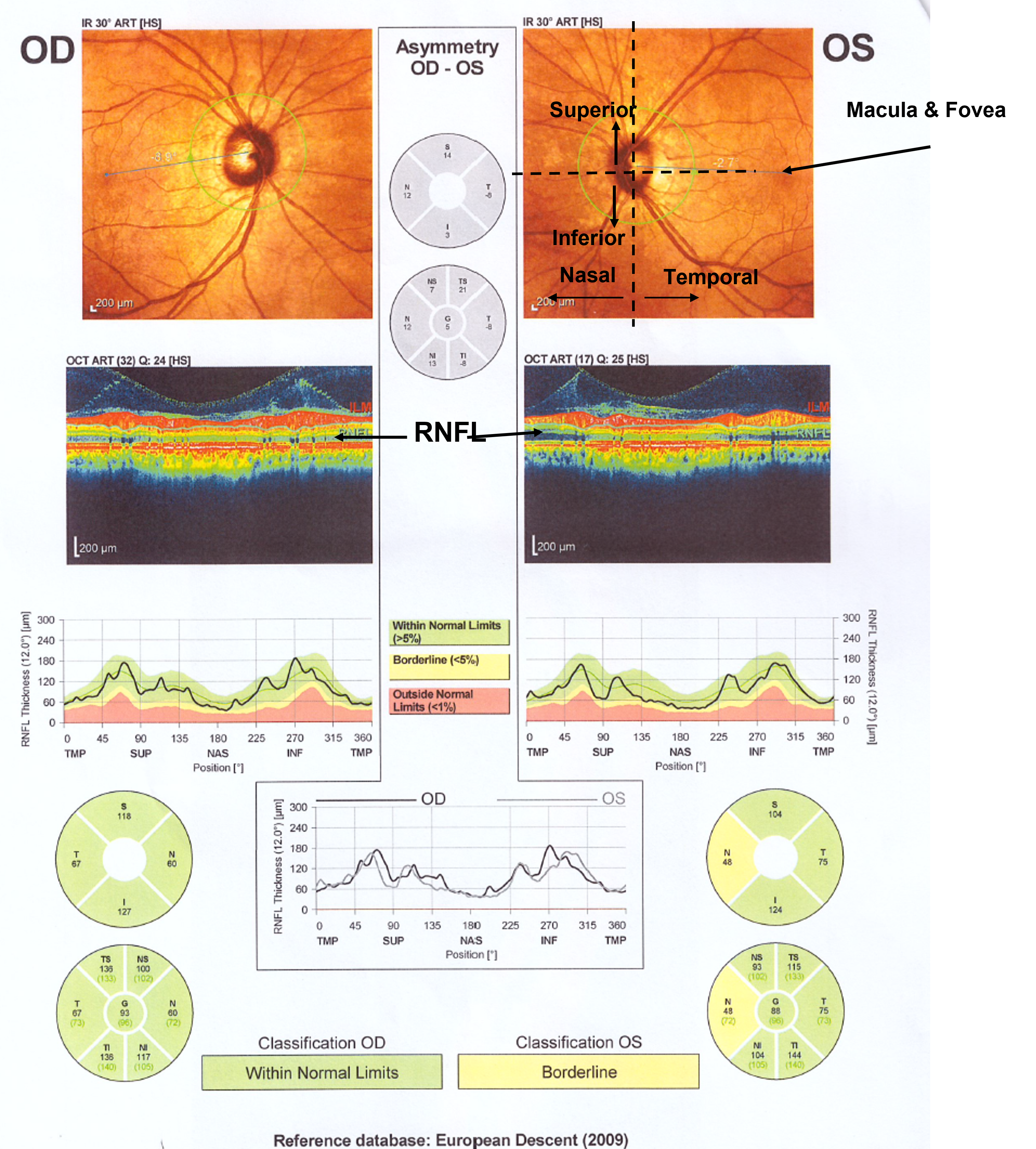}
\caption{An example OCT output that shows borderline Retinal Fiber Layer Thickness decrease in a patient. OD (oculus dextrus) is the right eye, and OS (oculus sinister) is the left eye. The eye has a spherical shape. The vertical area closer to the nose is called Nasal Retina (NAS, N), and the other side is called Temporal Retina (TMP, T). The upper retina from the fovea is Superior (SUP, S), and the lower part is inferior (INF, I). Therefore, 2D visualization requires unfolding the circle to a rectangle (from TMP to SUP, then to NAS and back to TMP). The Retinal Nerve Fiber Layer (RNFL) shown in the middle image is automatically segmented by the OCT system; its thickness is measured and compared against the European Descent reference database. The results show that there is a borderline thinning on the right eye.}
\label{fig:oct}       
\end{figure}

\begin{itemize}
    \item The very first observation noted by most studies is the thinning of the retinal layers \citep{Song2021}. Several neurodegenerative diseases, such as AD, dementia, and Parkinson's disease, have a characteristic signature of Retinal Nerve Fiber Layer (RNFL) thinning. Figure \ref{fig:retina_oct} shows the layers of the retina. Retinal thinning often refers to the inner layers such as the RNFL (also called Stratum opticum), Ganglion Cell Layer (GCL), and Inner Plexiform Layer. These three layers are called ganglion cell complexes. Ganglion cells are vulnerable to neurodegeneration because of their related mitochondrial dysfunction and their unique composition of having axons without myelination \citep{Carelli2017}. Recent studies show that GCL is more likely associated with brain biomarkers \citep{Moran2022, Yuan2022}. Figure \ref{fig:oct} shows an OCT output for an anonymized healthy patient. The RNFL thickness values in both eyes are measured by the device and compared against the reference database. OD is the left eye, and OS is the right eye. The eye and the retina have a spherical shape. Therefore, 2D slice visualization requires a circle to be mapped to a rectangle.
    \item Secondly highlighted observations are reduced macular volume and thickness. The macula provides sharp, clear, straight-ahead vision. It is responsible for central and color vision \citep{Bear2016}. Figure \ref{fig:oct} illustrates the macula region in the eye. The middle of the retina images and their graphs show the thickness of the macula.
    \item Some studies found reduced choroidal volume and thickness \citep{Lopez-De-Eguileta2020, Gharbiya2014,Jonas2016,Trebbastoni2017},  while others disagree with this finding \citep{Song2021, Bulut2018}. The choroid layer is the supplier of nutrients to the retina, and it also maintains the temperature and volume of the eye \citep{Bear2016}.
\end{itemize}

\subsubsection{Optical Coherence Tomography Angiography (OCTA)}
\label{OCTA}

\begin{figure}
\centering
\includegraphics[width=0.9\textwidth]{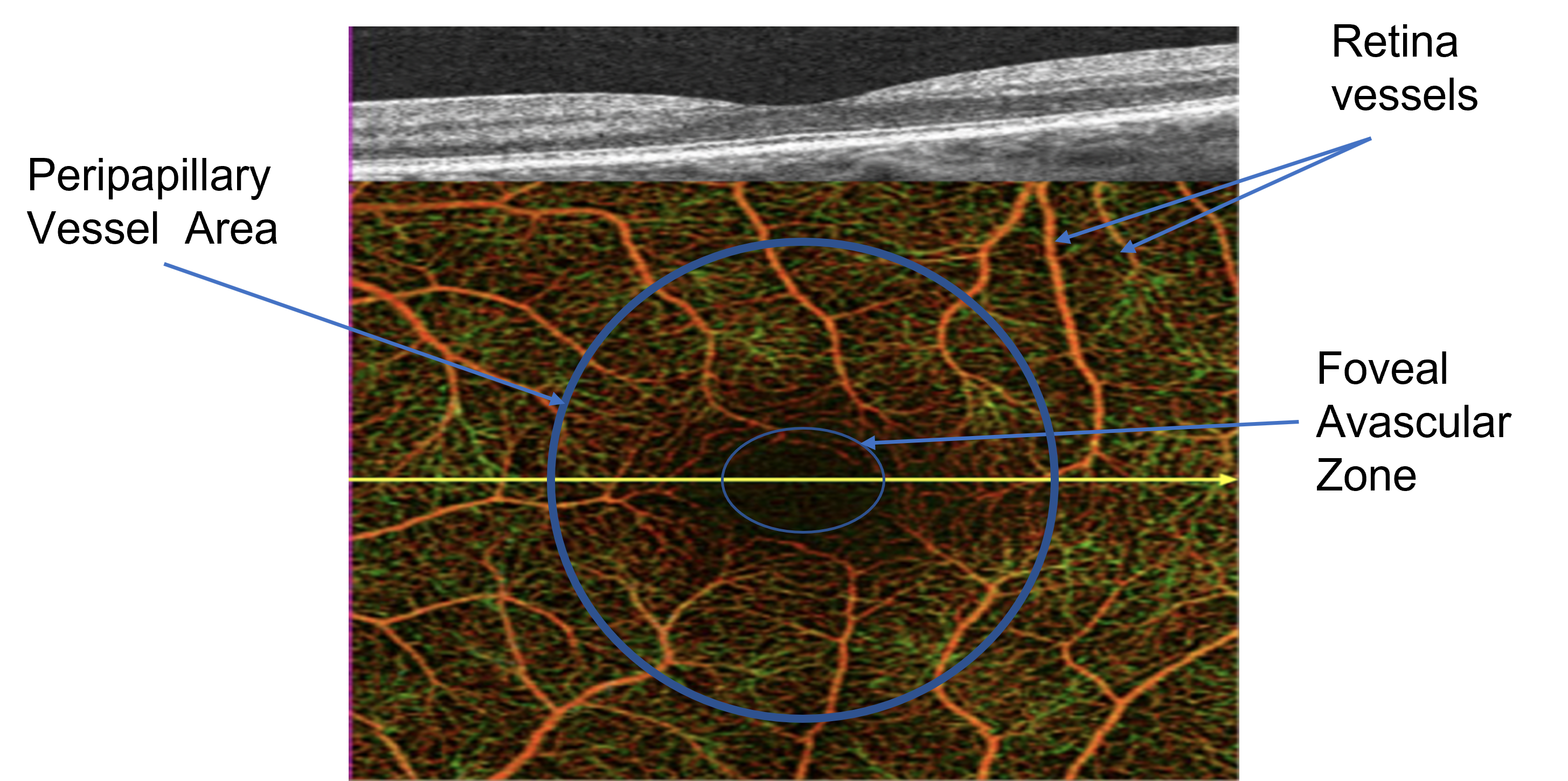}
\caption{OCT angiogram of a normal eye illustrating detailed microvasculature in the macula. The foveal avascular zone is seen in the center of the macula \citep{Green}. }
\label{fig:octa}       
\end{figure}

OCTA is a relatively new (2015) technique developed as a utility of OCT devices. Put simply, it is computed as the difference between two consecutive OCT scans to visualize the relative motion in the retina region, which reveals blood flow and vascular details at micrometer resolution. It offers a very effective imaging method in diagnosing eye diseases by unveiling microvascular changes or abnormalities in the blood flow pattern (Figure \ref{fig:octa}). Figure \ref{fig:octa} shows an OCT angiography of a healthy eye. Besides not being available in all OCT devices, it is susceptible to most OCT-induced distortions such as projection, motion, segmentation, and signal loss \citep{Attiku2021}.

In their survey, \cite{Song2021} also reviewed AD-related OCTA studies. Their major findings can be listed as follows:

\begin{itemize}
    \item Reduced vessel density is reported in several studies  \citep{Ge2021,Augustin2022}.
    \item Reduced perfusion density is also observed together with reduced vessel density. Vessel perfusion density is defined as the total area of perfused vasculature per unit area \citep{Triolo2017,Augustin2022}.
    \item Increased Foveal Avascular Zone (FAZ) is reported by \cite{OBryhim2021}. The fovea can be seen in Figure \ref{fig:oct} as the dark spot in the center where it marks the center of the retina \citep{Bear2016,Augustin2022}.
    \item Reduced peripapillary vessel density, another parameter that requires further research, is discussed in \cite{Ferrari2017}. The radial peripapillary capillaries are a distinctive vascular network within the RNFL around the optic disc, the area inside the circle in Figure \ref{fig:octa}.
\end{itemize}

\section{Machine Learning in OCT/OCTA}
\label{sec:Technical}

Newly introduced medical tests and imaging techniques help physicians diagnose diseases better. However, boosted by the advances in new imaging modalities, every patient has become a ``big data''
challenge \citep{Obermeyer2017} with a history of large-scale, high-dimensional, multimodal medical data. Diagnostic methods from this data greatly depend on physicians' professional experience and knowledge. Computer-aided diagnosis (CAD) have improved the performance of many challenging tasks, especially when working with high-resolution, multimodal, complex imaging data. Most of these systems have evolved from rule-based expert systems to fully automated and skilled deep learning systems now being used as medical decision support.

\begin{figure}
\centering
\includegraphics[width=0.8\textwidth]{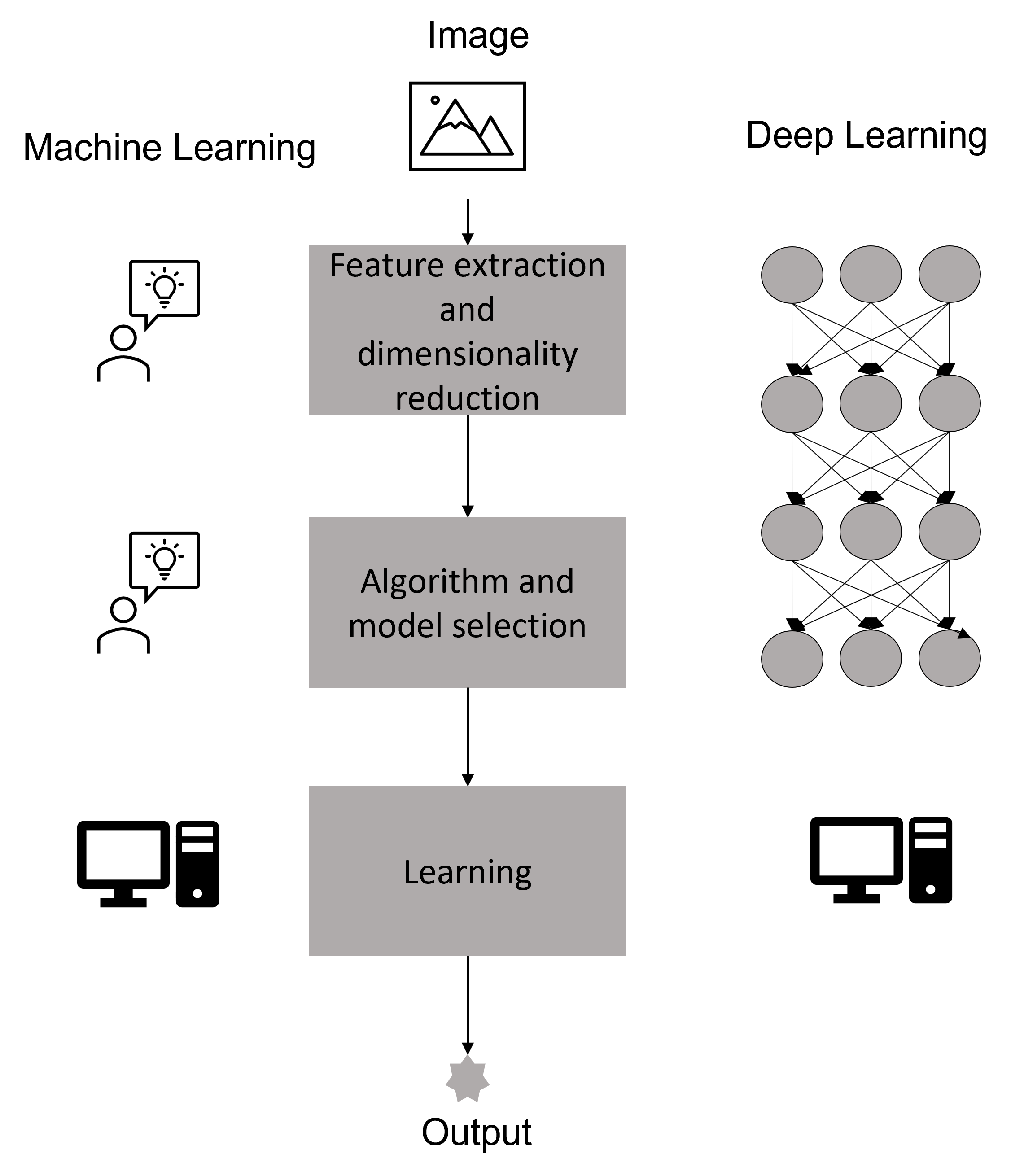}
\caption{Machine Learning techniques require expert intervention for features and the model. Deep Learning Networks learn feature extraction, representation and decision from the data.}
\label{fig:ml_dl}       
\end{figure}

Classical machine learning techniques require four common steps: feature extraction, dimensionality reduction, algorithm/model selection, and feature-based learning (Figure \ref{fig:ml_dl}). 
Artificial neural networks are a subset of AI inspired by a simplification of neurons and their connections in the brain. Deep learning (DL) is a multi-layer structure of neural networks that learns the features and high-dimensional non-linear functions by mimicking the neuron connections and interactions in the human brain. After the 2010s, DL gained momentum from Machine Learning \citep{Lecun2015} because it can learn from raw data by implicitly learning feature extraction, representation, and classification decision.

\begin{figure*}
\centering
\includegraphics[width=1\textwidth]{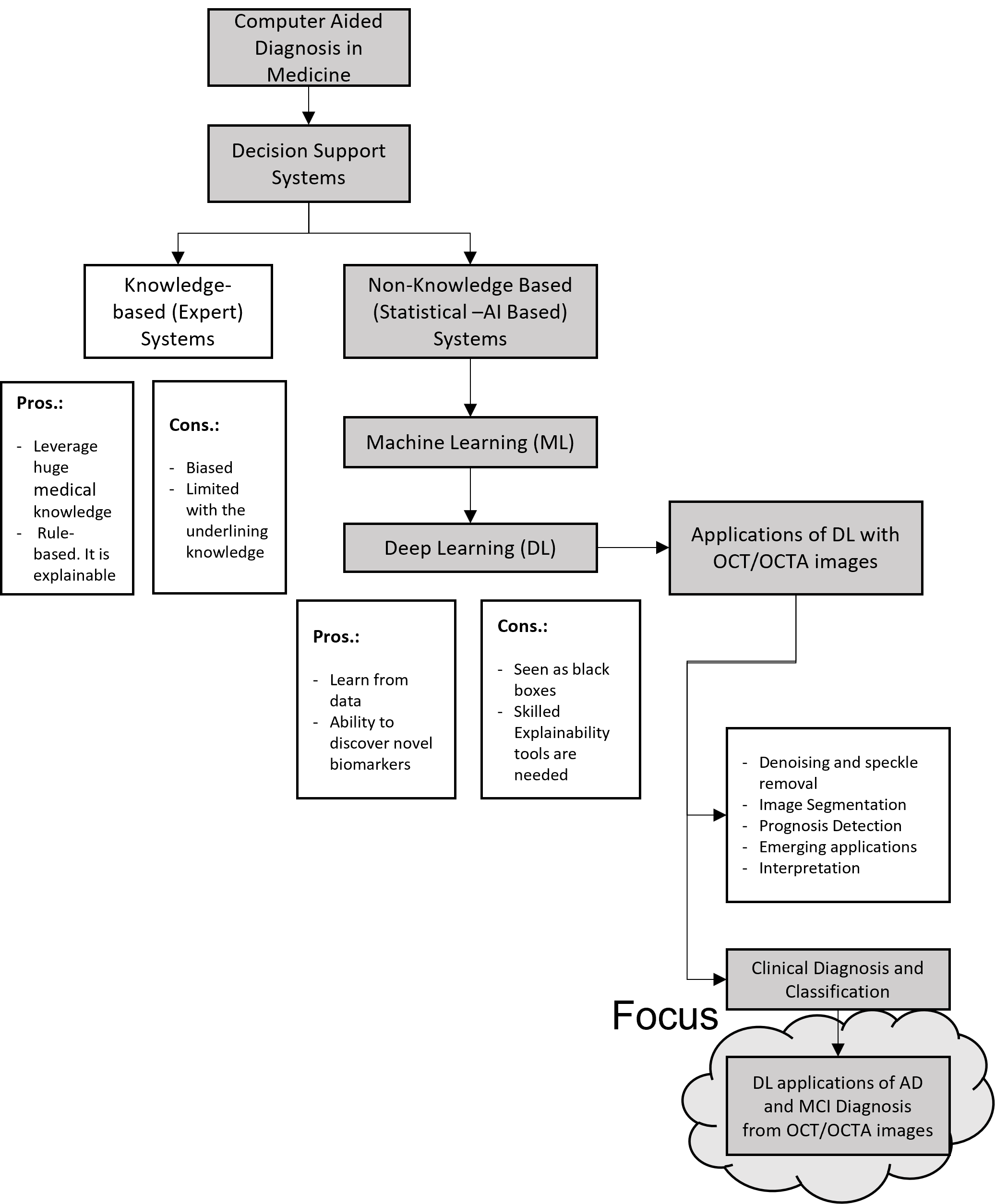}
\caption{Deep Learning is one the most powerful emerging technologies in computer aided diagnosis in medicine. Therefore, the main focus of the study is DL applications of AD and MCI Diagnosis from OCT/OCTA images. }
\label{fig:technical_flow}       
\end{figure*}

\begin{figure}
\centering
\includegraphics[width=0.8\textwidth]{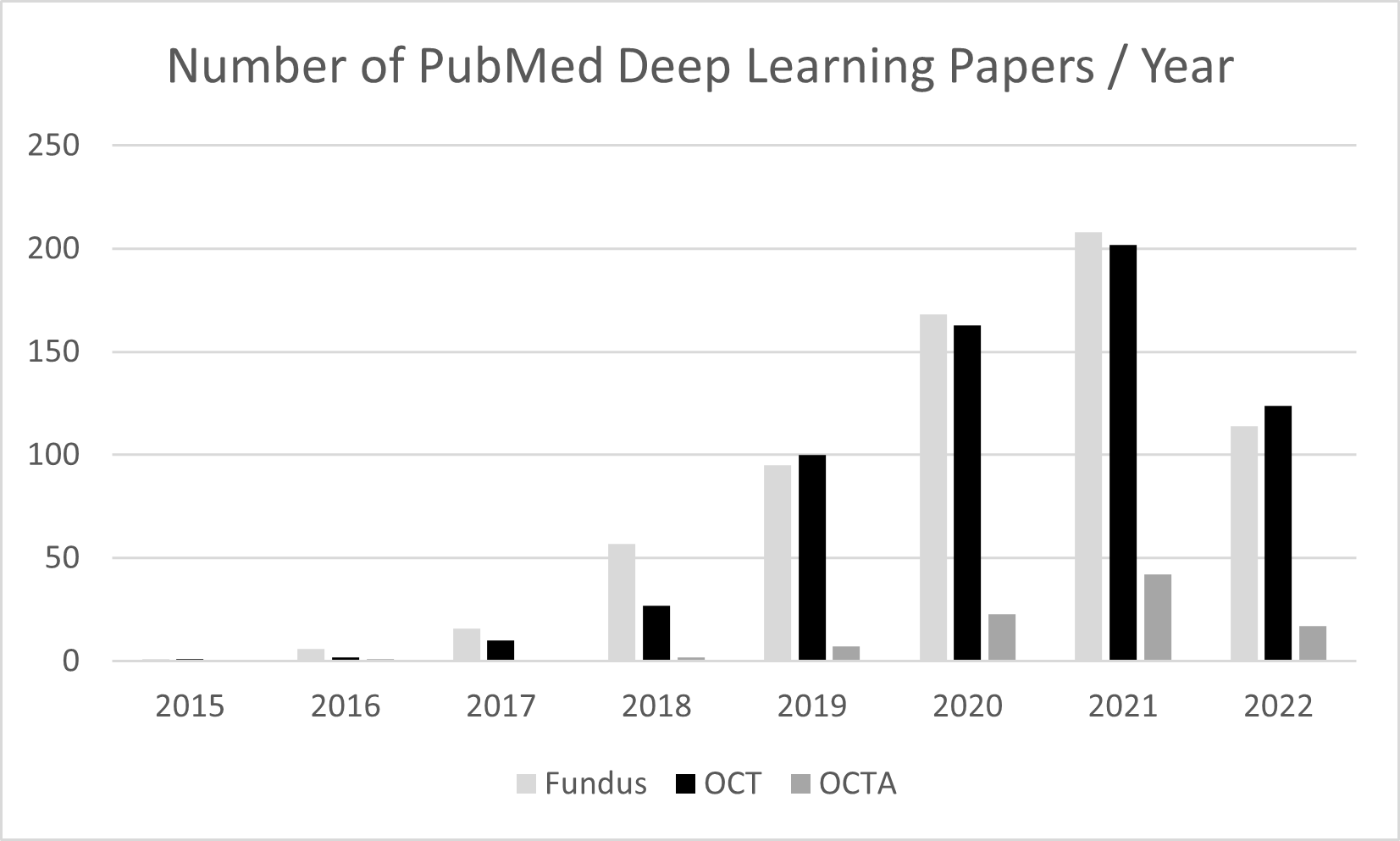}
\caption{PubMed statistics of DL Fundus, OCT and OCTA applications in years (2022 data covers only the first 7 months of the year).}
\label{fig:pubmed}       
\end{figure}

\cite{Tian2021a} reviewed application of the DL-based approach to different biomedical optical imaging fields such as microscopy, fluorescence lifetime imaging, in vivo microscopy, widefield endoscopy, photoacoustic imaging and sensing, diffuse tomography, and functional optical brain imaging. Unsurprisingly, DL has also become the latest trend in the analysis of OCT and OCTA scans \citep{Tong2020,Ran2021a,Latha2021}. Figure \ref{fig:pubmed} shows the number of DL studies on Fundus, OCT and OCTA images reported by PubMed in the last six years. The limited number of publications related to OCTA reflects its recent development and late availability in OCT devices.

The following are the major applications of deep learning in OCT/OCTA modality \citep{Tong2020,Tian2021a,Ran2021a}:
\begin{itemize}
    \item \textbf{Denoising and Speckle Removal:} Image quality improvement using computer vision (CV) and ML is one of the hottest topics in medical imaging. Due to the nature of the clinical practice, there are many issues concerning the scanned images. Raw OCT slice images suffer from speckle noise because of coherent light scattering (Figure \ref{fig:biobank}). Some devices have other noise sources due to low signal levels. Therefore, DL-based image quality improvements are investigated in many OCT/OCTA studies \citep{Tian2021a,Ran2021a,Huang2021}. Retina scans for each patient are taken at different times. Therefore, the datasets contain multiple images taken from various sensors at different periods. Besides, the retina is a non-rigid moving organ inside a moving body. Therefore, image registration may also be needed to match multiple images before the segmentation \citep{Pan2021} or classification processes.   
    
    \item \textbf{Image Segmentation:} OCT devices reveal structural details of the retina (see Section \ref{OCT}). Due to the layered structure of the retina, segmentation of the layers is the most common problem addressed by DL studies. Among all of the DL models, U-Net is the most popular one. RNFL and GCL layer thicknesses are the critical parameters for Alzheimer's studies. New OCT devices can measure RNFL thickness \citep{Schott2020}. However, other parameters such as the macula and choroid layer size require further research before manufacturers incorporate these options into future devices. Blood vessel segmentation is another parameter used to diagnose Alzheimer's disease. Even though there is extensive research on Fundus Photography, OCTA is now a trending technique that gives more information with higher-quality images \citep{Fan2021}. FAZ area segmentation is another new research topic in DL \citep{Mirshahi2021,Ran2021}.
    
    \item \textbf{Clinical Diagnosis and Classification:} OCT images are used to detect various diseases such as Age-related Macular Degeneration (AMD), glaucoma, diabetic macular edema (DME), diabetic retinopathy, retinal vein occlusion (RVO), choroidal neovascularization (CNV)  \citep{Latha2021,Zhang2014,Ran2021}. \cite{Latha2021} have reviewed studies on detecting ophthalmic disorders AMD, glaucoma, and diabetic retinopathy using DL techniques. \cite{Ran2021} did an extensive survey on DL techniques used to detect glaucoma. All of these studies are on single eye diseases such as glaucoma, AMD, etc. Eliminating multiple disease patients in these studies is a common practice that is not realistic from the diagnostic perspective. \cite{Ran2021a} reviewed the newer studies on deep learning-based multiple retinal disease detection on OCT/OCTA.
    
    \item \textbf{Prognosis Detection:} Treatment response and progress need to be assessed for diseases such as DME, CNV and RVO. OCT scans are used to track the changes in response to intravitreal anti-VEGF (vascular endothelial growth factor) agents in these diseases \citep{Ran2021}.  
    
    \item \textbf{Explainability:} Despite their high performance, deep neural network architectures are black box models. Trusting their predictions is essential in using them for decision-making in medicine. \cite{Ribeiro2016} raised the question ``Why Should I Trust You?'' and argued that people would trust a model more if they could understand the reasoning behind its predictions. The most common interpretability method used in OCT/OCTA, following other fields such as CT and MRI, is the GradCAM method \citep{Choi2021,Chueh2021,Yan2021}. \cite{Singh2021} evaluated 13 explainability deep learning methods for OCT scans. They assessed the Inception-v3 model that diagnoses three different diseases: CNV, DME and drusen, and found that the Deep Taylor (DTaylor) method \citep{Montavon2017}  outperforms the other techniques in OCT retinal images.
    
    \item \textbf{Emerging Applications:} \cite{Tian2021a} reviewed some emerging imaging modalities such as swept-source OCT \citep{Eid2022,Mopuru2022} and Doppler OCT (DOCT) \citep{Wada2020} that measure microangiography and blood flow. These studies gained momentum (see Figure \ref{fig:pubmed}) in the last couple of years because they are also generating additional functional and vascular features on top of anatomical features.  
\end{itemize}

\cite{Bourkhime2022} recently presented a brief systematic review on machine learning for AD screening using novel ophthalmologic biomarkers. They found 13 ML-based studies, including various data types, imaging modalities, and ML techniques. They identified only two studies (in PubMed and Scopus databases) directly using OCT image data \citep{Nunes2019,Lemmens2020}. We will analyze those studies in detail.
In the following section, we present our systematic review which focuses specifically on the diagnosis of Alzheimer's Disease and Mild Cognitive Impairment from OCT and OCTA scans. We further elaborate on the related datasets.
\section{Review}
\label{sec:methods}

This survey follows the guidelines of the Preferred Reporting Items for Systemic Review and Meta-Analysis (PRISMA) \citep{Page2021} to review ML/DL-based approaches for AD or MCI diagnosis in OCT and/or OCTA scans.
Figure \ref{fig:Prisma} shows the PRISMA flow chart of our systemic review process.

\textbf{Information Sources:} We performed an exhaustive search of English-language studies from Pubmed, Web of Science, Scopus, Google Scholar, Semantic Scholar and CrossRef from 01/01/2015 to 07/31/2022.

\textbf{Search Strategy:} We surveyed the databases using a software program called Publish or Perish \citep{Harzing2010} which retrieves and analyzes academic citations. Our search strategy was based on the following combination of terms: 

\textit{(``Alzheimer's'' OR ``dementia'' OR ``cognitive impairment'') AND  (``optical coherence tomography'' OR ``optical coherence tomography angiography'' OR ``Retinal Imaging'') AND  (``Machine Learning'' OR ``Deep Learning'')}

\textbf{Eligibility criteria:} Articles passing all the selection criteria were collected. The articles were excluded if they:
\begin{itemize}
    \item Are not available in English
    \item Are not published as a primary research paper in a peer-reviewed journal and are conference papers only;
    \item Do not describe the ML model for AD detection, screening or prediction using OCT/OCTA scan images or derived data.
    \item Are duplicates, datasets or book chapters
    \item Include other diseases apart from AD (such as AMD, Drusen, etc.)
    \item Are only related to segmentation or image quality improvement.
    \item Study other modalities apart from OCT and OCTA scans.
    \item Include only statistical data analysis.
\end{itemize}

\textbf{Data extraction, synthesis and analysis:} We analyzed the eligible studies and extracted information such as the number of participants, year of publication, algorithms applied and their characteristics, model prediction parameters, and model performance, including accuracy, discrimination sensitivity, and specificity rates. The analysis results are explained in the following section.
\subsection{Results}
\label{sec:Results}

\begin{figure*}
\centering
\includegraphics[width=0.9\textwidth]{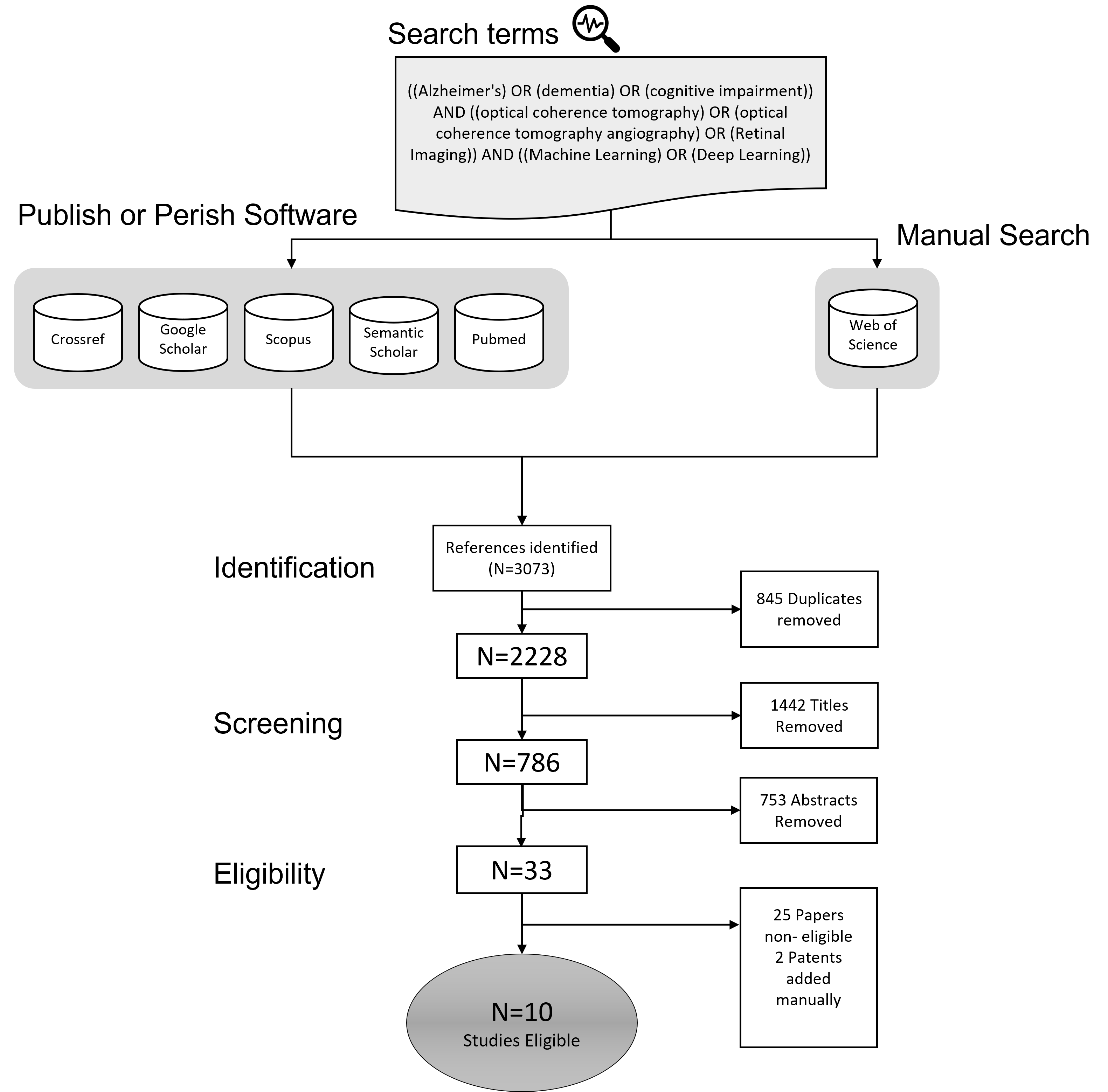}
\caption{The flow chart of the review results: ML/DL-based studies aiming at AD or MCI diagnosis using OCT and/or OCTA.}
\label{fig:Prisma}
\end{figure*}

The initial search returned 3073 references. After several steps of elimination (Figure \ref{fig:Prisma}), only ten were found eligible, including the two research papers already identified by \cite{Bourkhime2022}. Table \ref{tbl:Results} lists all the publications that we have reviewed in detail.

\cite{Nunes2019} followed a classical machine learning flow to discriminate between healthy (NC), Alzheimer's (AD) and Parkinson's disease (PD) subjects. First, they segmented OCT volumes to extract retina layers, then they computed the mean fundus image for the six retina layers (RNFL, GCL, IPL, INL, OPL and ONL) \citep{Guimaraes2014}. They computed the thickness of layers. Moreover, they computed Gray Level Co-occurrence Matrix and wavelet transform to extract textural features based on the layer images, resulting in 86 measurements per layer. The whole feature vector was classified using a support vector classifier. They reported 88.7\%, 79.5\% and 77.8\% median sensitivities for NC, AD and PD classes. Their findings showed that the textural features calculated in different retina layers may provide valuable information that is independent of layer thickness measures.

\cite{Sandeep2019} used a Fixed-Grid Wavenet Network (FGWN) \citep{CS2017} for the segmentation of OCT scans to identify the RNFL layer. Then they extracted morphological features such as area, perimeter, curvature variance, ellipticity from the RNFL image. They compared the back-propagation (BP) and radial basis function (RBF) neural networks for the classification and concluded that the RBF (reported in terms of false acceptance and false rejection rates). Their study concluded that the segmentation algorithm played an important role in the automated diagnosis of AD from OCT images.

\cite{Lemmens2020} combined snapshots of hyperspectral images with OCT data to identify AD patients among 10 probable AD, 7 biomarkers-proven AD and 22 NC subjects. Using hyperspectral images, they computed average reflectance from four different regions and combined this with RNFL thickness in four different eye regions (superior, inferior, temporal and nasal) computed from the OCT. They reported an average area under the curve of 74\% for the nested leave-one-out classification accuracy. 

Four other works experimented on animal subjects. The first study by \cite{Bernardes2017} used 4 and 8 months old 23 triple transgenic mice with AD (3xTg-AD) and  22 wild-type (WT) control mice. They segmented OCT scans into three layers. The first layer contained RNFL, GCL, IPL, INL and OPL. The second layer included the ONL and ELM, while the ellipsoid zone and RPE layers formed the third layer. Then they calculated a vector of features (histogram of OCT images, mean value of fundus images, and energies and contrasts of OCT images from different angles). The classification was performed by radial basis kernel SVMs. They compared the performance for 4 and 8 months old mice and concluded that predictions reached an accuracy over 80\% for mice at the age of 4 months and over 90\% for those at the age of 8 months.

The second animal study \citep{Trindade} worked with 57 3xTg-AD and 58 WT mice Healthy Cohort (HC). They segmented OCT data to compute six different images corresponding to the following layers/layer-aggregates: RNFL-GCL, IPL, INL, OPL, ONL, and the total retina (TR), which encompasses all anatomical layers. They implemented six convolutional deep learning models (CNN), Inception-v3 to learn and classify fundus images computed from six different layered AD and HC OCT scans. They reported that RNFL-GCL achieved the highest classification accuracy (89.2\%). 
Feeding the output of six CNNs into a feed-forward neural network (FFNN) to combine information from different layers did not improve the classification performance further. The study also used GradCAM explanations to produce heat maps of the most effective areas for classification. Moreover, they noted that the right eye and the left eye may convey different information. 

The third study with mice was performed by \cite{Sayeed2022} with 24 3xTg-AD and 2256 WT mice. Their study tested various ML methods such as decision trees, neural networks, random forest and SVM. They generated 22 different features (e.g. contrast, sum entropy, difference entropy) from a normalized covariance matrix of Fundus and OCT images. SVM has the highest performance results (Accuracy 99\%, Sensitivity 98\% and Specificity 99\%). The performance of the decision tree (Accuracy 96\%, Sensitivity 94\% and Specificity 97\%) was the second best, while the neural network achieved the worst performance (Accuracy 74\%, Sensitivity 98\%, Specificity 73\%).

\cite{Ferreira2022} performed the latest study with mice using 57 3xTg-AD and 57 WT mice at different ages. They generated mean-value-fundus (MVF) images from segmented OCT layers RNFL-GCL, IPL, INL, OPL and ONL. The images were pre-processed to increase the performance, and a data augmentation technique was used to reduce over-fitting. Transfer learning from the inception-v3 model was used for training the network with mice at 3,4 and 8 months. For the classification tests, 1,2 and 12 months old mice were used. They concluded that the best performance was obtained using MVF data calculated using the RNFL-GCL (Accuracy 88.1\%, Sensitivity 87.7\%, Specificity 88.5\%, and F1 score 88.5\%). They also measured the error rates of mice at different ages. As expected, there were smaller classification errors with validation data than with the test data. They also concluded that the total classification error at 12-months-old (15\%) was smaller than  2-months-old (16.9\%) even though 2-months-old was much closer to the trained age group.  

A recent study performed by \cite{Wisely2022} employed a deep residual convolutional model (ResNet18) to detect AD using multimodal retinal images. They used image data (Color maps of GC-IPL thickness of OCTA, OCTA, ultra-widefield (UWF) color and fundus autofluorescence images), numerical data (such as RNFL thickness information) obtained from OCT/OCTA and patient information as simultaneous input to an artificial neural network (ANN) through separate channels. Two-dimensional images were fed into convolutional neural network branches, and OCT-based numerical information and patient information were processed in a fully connected subnet structure. They found GC-IPL to be the best performing input (AUC 0.809). GC-IPL was calculated over OCT like RNFL, reflecting a measurement superimposed on OCTA in the corresponding study. It was stated that the combination of numerical information obtained from OCT and GC-IPL images gave the highest classification performance (AUC 0.841). Non-OCT modalities such as FFA and UWF do not improve classification performance. The study is important in demonstrating the effectiveness of GC-IPL and OCTA (AUC 0.828). Though the study used quantitative data and patient information to support 284 eye images from 159 patients, the data size was still quite limited, which is the top known cause of overfitting in neural networks. The dataset was collected in a cross-sectional study (\citep{ClinicalTrials}, NTC03233646). 

Finally, our manual search yielded two patents. The first patent application by Carl Zeiss Meditec, Inc. (Patent No: US 6,988,995 B2) claimed a method based on artificial neural networks and RNFL Images to classify Alzheimer's disease. The second patent application by the National Chin-Yi University of Technology, Chi Medical Center (Patent No: US 11,288,801 B2) claimed CNN-based methods for both segmentation and classification of OCT images.

We could not find any research study on cognitive impairment with OCT/OCTA data using an ML model, even though there are two recent studies that use fundus images \citep{Corbin2022,Zhang2021}. However, we observed new data collection initiatives to include MCI cases (Section \ref{sec:dataset}).


\cite{Yanagihara2020} reviewed the issues concerning DL in OCT for retinal diseases. They highlighted the issues of limited number of datasets, lack of image collection standards, inconsistent metrics, and computational restrictions. Even though all of these challenges apply to all works, it seems as if the study of AD or MCI requires further consideration because we could identify only ten AD studies out of 3073 deep learning studies. The following section explains our search on related datasets.





\begin{table}
\centering
\caption{Development details of ML/DL models using OCT and OCTA data. GC-IPL: ganglion cell-inner flexiform layer, RNFL:Retinal Nerve Fibre Layer, IPL: Inner Plexiform Layer, INL: Inner Nuclear Layer, OPL: Outer Plexiform Layer, Sens: Sensitivity, Spec: Specificity, Acc: Accuracy, 3xTg-AD:triple transgenic AD, WD: wild-type\\ }
\label{tbl:Results}%

\resizebox{1\textwidth}{!}{%

\begin{tabular}{ | m{1.6cm}| m{1.5cm}| m{2.4cm}| m{2cm}| m{4cm}| m{2.3cm}| m{3cm}| m{7cm}| } 

\hline
Author, Year  & Country & Source Data & Study Design & Main Variables & ML Algorithms & Validation method, feature detection (FS), data prepossessing (DP)  & Performance \\
\hline
\hline
\cite{Ferreira2022} & Portugal	&  University of Coimbra &	57  3xTg-AD and 57 WT mice  &	Mean Value of Fundus Image &	Convolutional Neural Networks (Inception-v3)	 & Data split for train, test and validation, FS, DP	& Acc of 4 months=76\% Acc of 8 months=92\%
\\
\hline
\cite{Sayeed2022} & India & ACE College of Engineering, Trivandrum 695027, Kerala,& 24 3xTg-AD , 2256 WT mice  & Derived features from Fundus and OCT images & Decision Tree, Neural network, Random Forest, SVM & Validation method not specified, FS, DP & Decision Tree (96\% Acc., 94\% Sens, 97\% Spec)
Neural Network(74\% Acc., 98\% Sens, 73\% Spec)
Random Forest (98\% Acc., 65\% Sens, 98\% Spec)
SVM (99\% Acc., 98\% Sens, 99\% Spec) \\
\hline

\cite{Wisely2022} &	US & Duke University, NCT03233646 &	123 HC, 36 subjects with AD & Image Data from various modalities, patient data, derived quantitative OCT data &	Convolutional Neural Networks (ResNet18) &	186 eyes for training and validating, 68 eyes for testing, FS, DP &	GC-IPL maps AUC=0.809. All images, quantitative data and patient data AUC=0.836. All images AUC=0.829. GC-IPL maps, quantitative data and patient data AUC=0.841. \\
\hline
\cite{Trindade} &	Portugal	&  University of Coimbra &	57 3xTg-AD, 58 WT mice &	Fundus images + layer-aggregates RNFL-GCL, IPL, INL, OPL, and Total Retina(TR) &	Convolutional Neural Networks (Inception-v3) & five-fold cross validation &	6 CNNs trained separately resulted with Acc between 79.0\% and 89.2
\\
\hline
\cite{Lemmens2020} &	Belgium & not available	& 10 AD, 7 amyloid proven AD and 22 HC &	Amyloid accumulation and RNFL (retinal imaging hyperspectral snapshot camera + OCT) &	ML model not identified &	leave-one-out cross validation, FS, DP	& AUC of 0.74
\\
\hline
\cite{Bernardes2017} & Portugal	& University of Coimbra &	22  3xTg-AD and 23 WT mice  &	Histograms of OCT scans, Mean Value of Fundus Image and Energy and Contract values of OCT layers &	SVM	 & k-fold cross-validation, FS, DP	& Acc of 4 months=76\% Acc of 8 months=92\%
\\
\hline
\cite{Sandeep2019}	& India	& Department of ECE, College of Engineering, Trivandrum, Kerala &	25 AD, 25 HC &	Morphological Features from segmented RNFL  &	Fixed-Grid Wavenet Network, Radial Basis Function of Neural Networks	 & N/A, FS, DP	& False Acceptance Rate = 1\%, False Rejection Rate=1\%\\

\hline
\cite{Nunes2019}	& Portugal	& University of Coimbra &	20 AD, 28 PD, and 27 age-matched HC &	Retinal texture and thickness (fundus images by OCT) Demographic data &	SVM	 & k-fold cross-validation, FS, DP	& Sensitivity=88.7\%,
Specificity=84.9\%, Accuracy=82.2\%\\
\hline
\cite{Et2006}	& US &	Patent No:6,988,995 B2 &	N/A	 & RNFL	& Artificial Neural Networks	& not available	& not available \\
\hline
\cite{Zhou2010}	& US &	Patent No:11,288,801 B2
 &	N/A	 & RNFL	& Convolutional Neural Networks	& not available	& not available \\

\hline

\end{tabular}%
}
\end{table}

\begin{sidewaystable}
    \centering
    \caption{Alzheimer's disease and OCT data collection initiatives based on ClinicalTrials.gov\\}
    \resizebox{1\textwidth}{!}{
    \begin{tabular}{|l|l|l|l|}
    \hline
        Study Title & Conditions & Interventions & Locations \\ \hline \hline
        1. Finding Retinal Biomarkers in Alzheimer's Disease & Alzheimer Disease & Other: Detailed ophthalmologic examination & Centre Mémoire de Ressources et \\ \hline
        ~ & ~ & ~ &  de Recherche Paris Nord, Paris, France \\ 
        2. Relationship Between Alzheimer Disease and  & Optical Coherence Tomography & Optical coherence tomography & CHU Amiens \\
        Diminution of the Three Macular Nervous  & Optical Coherence Tomography Ang. &  Optical coherence tomography  & Amiens, France \\
    Retinal Layers & Retinal Thickening & angiography (OCTA) & ~ \\
        ~ & (and 5 more...) & ~ & ~ \\ \hline
        3. OCT Angiography and NRAI in Dementia & Alzheimer Disease & Optical Coherence Tomography Ang. & Oregon Health \& Science University \\ 
        ~ & Dementia &  (OCTA) Imaging & Portland, Oregon, United States \\ 
        ~ & Mild Cognitive Impairment & Noninvasive Retinal Amyloid Imaging (NRAI) & ~ \\ \hline
        4. The Role of Ophthalmologic Tests and EEG Imaging  & Mild Cognitive Impairment & Diagnostic Test: observational & Zhongshan Ophthalmic Center \\ 
        in Alzheimer's Disease & ~ & ~ & Guangzhou, Guangdong, China \\ \hline
        5. OCTA in Mild Cognitive Impairment and & Alzheimer Disease & Device: Retinal Imaging & Duke University Medical Center \\ 
         Alzheimer's Disease & Mild Cognitive Impairment & ~ & Durham, North Carolina, United States \\ 
        ~ & Retinal Vascular & ~ & ~ \\ 
        ~ & (and 4 more...) & ~ & ~ \\ \hline
        6. OCT-Angiography and Adaptive Optics in Patients  & Amnesia & Procedure: Ophthalmological exam & Hôpital Fondation A. de Rothschild \\ 
        With Memory Impairment & Alzheimer Disease & Procedure: Blood pressure measurement & Paris, France \\ 
        ~ & Lewy Body Disease & ~ & ~ \\
        ~ & Parkinsons Disease With Dementia & ~ & ~ \\ \hline
        7. Retinal Neuro-vascular Coupling in Patients  & Mild Cognitive Impairment & Device: DVA & Department of Clinical Pharmacology,  \\ 
        With Neurodegenerative Disease & Alzheimer Disease & Device: FDOCT & Medical University of Vienna \\
        ~ & Healthy & Device: Pattern ERG & Vienna, Austria \\
        ~ & ~ & Device: Optical Coherence Tomography & ~ \\ \hline
        8. BEAM: Brain-Eye Amyloid Memory Study & Alzheimer's Disease & Other: Pittsburgh Compound B [11C]-PIB & Sunnybrook Health Sciences Centre \\ 
        ~ & Mild Cognitive Impairment & ~ & Toronto, Ontario, Canada \\ 
        ~ & Vascular Cognitive Impairment & ~ & St. Michael's Hospital \\
        ~ & (and 2 more...) & ~ & Toronto, Ontario, Canada \\ 
        ~ & ~ & ~ & University Health Network \\
        ~ & ~ & ~ & Toronto, Ontario, Canada \\ 
        ~ & ~ & ~ & (and 2 more...) \\ \hline
        9. Atlas of Retinal Imaging in Alzheimer's Study & Alzheimer Disease & Other: Retinal Imaging & Morton Plant Hospital \\ 
        ~ & Mild Cognitive Impairment & Other: Pupillometry & Clearwater, Florida, United States \\ 
        ~ & Mild Dementia & Other: Contrast Sensitivity & St. Anthony's Hospital \\ 
        ~ & (and 2 more...) & (and 5 more...) & Saint Petersburg, Florida, United States \\ 
        ~ & ~ & ~ & University of Rhode Island \\ 
        ~ & ~ & ~ & Kingston, Rhode Island, United States \\
        ~ & ~ & ~ & Butler Hospital \\ 
        ~ & ~ & ~ & Providence, Rhode Island, United States \\ \hline
    \end{tabular}
  }
  \label{tbl:clinical_trials}

\end{sidewaystable}

\subsection{OCT/OCTA Datasets}
\label{sec:dataset}
In a 2020 review, \cite{Yanagihara2020} identified the lack of publicly available OCT/OCTA datasets as a significant obstacle for developing deep learning-based approaches to retinal diseases.
During our survey, we observed that the situation has not improved much. Most clinics collect local datasets with a limited number of patients for their own research. Basically, the AD-OCT data collection attempts remain local and private \citep{Bulut2016,Bulut2018,Bayhan2015,Nunes2019,Querques2019,Tian2021a,Ma2021,Lemmens2020}. This is perhaps due to the fact that collecting AD-related data requires substantial effort:
\begin{itemize}
    \item A recognized Ethics Committee should approve any medical data collection involving human subjects, and analysis should apply the principles stated in the ``World Medical Association Declaration of Helsinki: Ethical principles for medical research involving human subjects'' \citep{helsinki2013}. 
    \item All patients or their legal representative (for AD patients who cannot sign for themselves) should sign a written consent.
    \item All subjects should undergo clinical neurologic and neuropsychologic assessment (see Section \ref{AD_Diagnosis}) to determine their AD level. Patients with AD are excluded if there is a history or evidence of other neurologic or psychiatric disorders or other types of dementia.
    \item Especially in OCT/OCTA studies, ophthalmologists should exclude patients with other diseases such as diabetes mellitus, systemic arterial hypertension, cardiovascular diseases, retinal diseases (e.g. AMD), glaucoma, ocular disorders (trauma, surgery, inflammation). Some studies even exclude smoking \citep{Bayhan2015}.
    \item Cognitively healthy (HC) subjects with ages matched to AD subjects should be recruited for the study. Usually, AD subjects are over 65-70 years old. 
    \item Either issues concerning inconsistent and low-quality images should be resolved or the images with unresolved issues should be excluded. OCT is known to produce various image artefacts due to factors such as movement and signal loss. 
    \item Medical images should be read, assessed and marked by experts. Most of the time, the results need to be verified by more than one expert.
\end{itemize}
    
Data collection usually extends over several years. After the data is collected it is usually kept local as  there are strict data privacy rules that impede data sharing among researchers. However, publicly available medical data is priceless for academic research. For example, the Alzheimer’s Disease Neuroimaging Initiative (ADNI) is a longitudinal multicenter study designed to develop clinical, imaging, genetic and biochemical biomarkers for the early detection and tracking of AD \citep{ADNI}. This extensive dataset, collected in various phases and collaborations, is open to public access for researchers. Unfortunately, opthalmology has no global public dataset similar to ADNI. \cite{Khan2021} did extensive research on public datasets that contain ophthalmologic health information. \cite{Bissig2020} published the only publicly available AD dataset (14 AD and NC patients) on the Dryad site \citep{Bissig}. They studied light-dependency changes in retinal reflectivity on glial cell function as a new biomarker for AD. The images were taken either in the dark or bright, stimulating light. Therefore, they were not consistent with standard AD datasets. There are other local, non-AD data collection initiatives such as Retinal OCTA Vessel Segmentation Dataset (ROSE) \citep{Ma2021} and Annotated Retinal OCT Images (AROI) database \citep{Melinscak2021}.

Alzheimer's Association in the United States supports a collective initiative called the Atlas of Retinal Imaging in Alzheimer's Study (ARIAS) \citep{Alber2020}. In this study they defined a ``minimum dataset'' framework for SD-OCT retinal image acquisition and analysis. Then, there is an ongoing data collection initiative in accordance with this framework  \citep{Alber2020b} (See Table \ref{tbl:clinical_trials}). The dataset was planned to be a publicly available longitudinal study of structural, functional, metabolic and angiographic retinal AD biomarkers in a large cohort (N=330 aged 55-80; 50 NC–low AD risk, 200 NC–high AD risk, 50 MCI, and 30 mild AD). There are other initiatives for standardizing retinal OCTA imaging as discussed in the review by \cite{Sampson2022}. 

\begin{figure}
\centering
\includegraphics[width=0.7\textwidth]{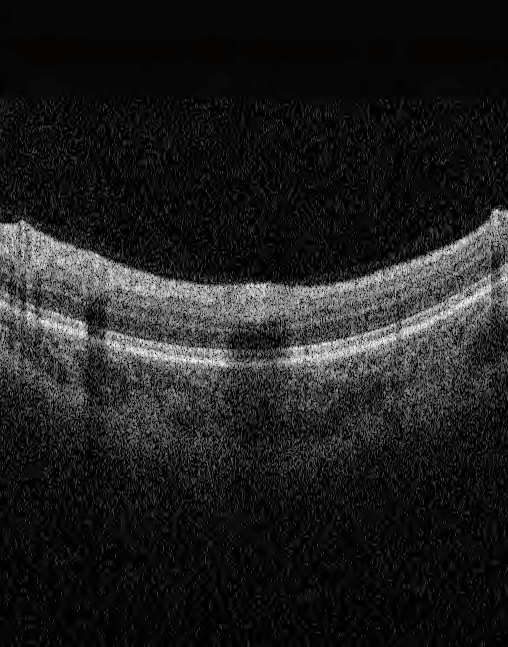}
\caption{One of 128 OCT scan slices of a sample UK Biobank data. The raw data has speckles and noise that need to be cleaned before using in any study.}
\label{fig:biobank}       
\end{figure}

UK Biobank \citep{UK_Biobank,Chua2019} offers an extensive biomedical database containing genetic and health data collected from half a million UK participants. The database is open to researchers, but at a cost. This database includes all the clinical examinations, physical, mental and ophthalmologic data with demographic information, OCT scans, derived OCT, and AD measures for each patient. Therefore, it is a complete resource for studying AD-related changes in the eye. OCT scans of 84,778 UK patients (both eyes) were acquired by TOPCON 3D OCT 1000 Mk 2 device and stored in TOPCON (.fda and .fds) and image (.png) formats. Figure \ref{fig:biobank} shows one of 128 OCT slices from the UK Biobank dataset.  \cite{Tian2021} used fundus images and related data from this dataset to classify AD from the retinal vasculature.

AlzEye is another major UK-based data collection initiative run by Moorfield Eye Hospital NHS Foundation Trust \citep{Wagner2022}. The data was collected from 353,157 participants between 2008 and 2018. This data collection is the most extensive ophthalmology data for  Alzheimer's disease. More than six million retinal images of seven different modalities (Color Fundus, Red-free photograph, fundus autofluorescence (widefield), pseudocolor photography (widefield), OCT and OCTA) were collected using devices from three different vendors (Topcon, Heidelberg and Optos). The dataset is not public yet. 

There are newer OCT/OCTA dataset collection initiatives concerning  Mild Cognitive Impairment and Alzheimer's Disease patients \citep{Yoon2019}. Our \cite{ClinicalTrials} database search returned ten other ongoing data collection initiatives on Optical Coherence Tomography and Alzheimer's Disease, summarized in Table \ref{tbl:clinical_trials}.

Generating synthetic data by learning previously collected AD datasets is sometimes useful in improving the accuracy of classifiers. Even though this may increase the AD sample size, it may lead to a sample bias. \cite{Danesh2021} have devised a method called Active Shape Modeling (ASM) \citep{Cootes1995} to generate 3D OCT images with similar anomalies to expand their training datasets.

\section{Discussion}
\label{sec:discussion}

We surveyed and reviewed the ML or DL-based studies to detect or diagnose AD in OCT and OCTA scans. We aimed to investigate the medical and technical background of the problem thoroughly and reviewed existing technical proposals and datasets.

Our survey of the medical literature has revealed the most recent findings about AD-related changes in the retina, which are observable through OCT examination. The current literature focuses on layer thickness changes in particular retinal layers such as Retinal Nerve Fiber, Ganglion Cell, and Inner-Plexiform Layers. Usually, the OCT scan is segmented automatically (sometimes with manual interventions) to identify these layers; and then layer average thickness values are computed. Medical studies have revealed that the average thickness of those layers in AD and normal patients show statistically significant difference. 

Our survey of technical literature has revealed that the proposals for enhancing and analyzing the scans are growing exponentially. Currently, the ML or DL-based scene is populated by automatic segmentation studies or clinical diagnosis and classification of other diseases such as glaucoma and diabetic retinopathy. Some works based on neural networks already exploit explainability tools such as GradCAM \citep{Selvaraju2020}, DTaylor \citep{Montavon2017} in OCT and OCTA modalities. 

Our PRISMA-based \citep{Page2021} systematic review of the ML/DL approaches to AD or MCI diagnosis in OCT scans has extended the recent review of \cite{Bourkhime2022} on AD diagnosis using OCT-based biomarkers. We identified eight additional studies, four of which used mice subjects and two of which were US Patents. We observed that the scene from classical machine learning pipeline is moving towards neural networks-driven fully automated classification. In this transformation, we observed the use of specific layer thickness measurements and hand-crafted textural metrics on images as features (inputs). In addition, the sample sizes in the reviewed works were quite limited, usually containing less than 50 AD subjects. A full deep learning solution to process raw OCT scans as input has not been developed yet, probably because it would require substantially more AD subjects. As expected, the different methods are not yet comparable to each other because they all use different datasets and accuracy measures.

The study by \cite{Wisely2022} was the only work that used OCTA images and other data together in a neural network-based structure. Although they have tried other imaging modalities (e.g. ultra-widefield color and fundus autofluoresence) as well, they have found that color maps of Ganglion Cell Inner Plexiform Layer thickness overlayed on OCTA was the best input for AD vs. non-AD classification. Using OCT/OCTA numerical data (e.g. RNFL thickness, area of foveal avascular zone) and patient data (demographics and MMSE scores) together with GC-IPL was found to be the best input combination.

OCTA is a relatively new technique developed as a byproduct of OCT scans. It computes the difference between two consecutive OCT scans to reveal blood flow and micro-vascular retinal structure at micrometer resolution. So far, medical research has identified reduced vessel density in different regions and increased foveal zone in AD patients. AD-related research in both OCT and OCTA modalities seems to be just gaining momentum. Presumably, more biomarkers will be discovered with the increasing support expected from the technical side.

Deep Learning models are data hungry. Therefore, most of the ML/DL studies \citep{Wisely2022, Tian2021, Lee2017,Thakoor2022} have used both eyes (left and right) to double the dataset size. \cite{Nunes2019} found that when both eyes received the same classification, the classification accuracy improved from 82\% to 96\% in this group. 

Most relevant research used 2D OCTA images or OCT computed thickness values. Increasing computational power and cloud services allow 3D volumetric retinal images to be analyzed. Working with 3D OCT scans may also change the design of the problem and speed up the search for novel biomarkers. However, such a dramatic increase of input dimensionality will require substantially more data.

Publicly available data is a priceless resource for academic research.
\cite{Yanagihara2020} listed four challenges of deep learning approaches to OCT image analysis: 1) lack of large datasets, 2) non-standardized image acquisition and preprocessing, 3) limited computational power, 4) inconsistency of reporting metrics.  
Two years later, our survey on AD or MCI-related datasets has shown no significant improvements, which seems to be the main issue against the development of deep learning approaches. We found that UK Biobank offers its database to global researchers applying via a research proposal. Although access to the database requires a compensation fee, there are reduction schemes to support research students. AlzEye initiative by \cite{Wagner2022} seems to be currently the largest dataset for ophthalmic images. However, the dataset is not yet open to the public, although it may become available through collaboration. 

Nevertheless, AD opthalmology requires a large-scale global dataset such as Alzheimer’s Disease Neuroimaging Initiative (ADNI) \citep{ADNI}, which includes MRI, PET, and related data of thousands of subjects collected by the ADNI and their collaborative studies. The data will be the main obstacle impeding progress in the field until such a dataset is created. . 

Table \ref{tbl:clinical_trials} lists many local ongoing data collection initiatives which are promising. One of the significant efforts is by Alzheimer’s Association in the United States which supports a large-scale (16 instituions) longitudinal data collection study: the Atlas of Retinal Imaging in Alzheimer’s Study (ARIAS) \citep{Alber2020}. To support the initiative, they have defined a ``minimum dataset” framework to standardize SD-OCT retinal image acquisition and analysis. More initiatives using the minimum framework would help solve both the open dataset problem and standardization issues. 

Alzheimer's disease patients are aged people with more potential for multiple dementia \citep{Gaugler2022} and multiple retinal diseases such as age-related macular degeneration \citep{Zhang2022}. There were various studies on deep learning-based multiple retinal disease detection using OCT/OCTA \citep{Ran2021}. Unfortunately, none of them included AD or MCI. In new data collection initiatives, multi-disease AD samples could also be included in the datasets to allow differential diagnostic studies on AD diagnosis with multi-retinal diseases. 

Alzheimer's disease is a prolonged progressive disease. Therefore, detection of the disease in its early stages is vital. During our research, we observed that despite an increasing focus on early/mild cognitive impairment, there are no deep learning and/or machine learning studies on MCI yet. Once datasets such as \cite{Alber2020} become available, we can expect to see more studies on this topic.

\section{Conclusion}
\label{sec:conclusion}

 In this study, we surveyed the diagnosis of Alzheimer's Disease and Cognitive Impairment using OCT and OCTA imaging of the retina. Our survey included an introduction to the medical background of the diseases and a review of various technical proposals from an automated image analysis perspective. We reviewed recent (2015-2022) deep learning approaches and OCT/OCTA datasets directly aiming at AD and MCI diagnosis. We identified eight ML/DL studies and two patents aiming at AD diagnosis, which shows that the research field is still in the conception phase. Our survey on datasets identified the lack of open OCT/OCTA datasets about AD as the main issue impeding the progress in the field. We observed that there are many recent initiatives for OCT data collection aiming at AD diagnosis. Some important longitudinal dataset initiatives such as ARIAS are promising to fuel the progress in the field. In addition, there is an increasing focus on MCI and OCT in clinical studies since MCI can be an early cursor for AD. Even though there are no machine learning and/or deep learning studies on this topic yet, it will attract attention when the datasets become available. The current studies usually leave out data acquired from patients with other eye diseases or abnormalities that are frequently seen in the elderly population. The development of a realistic diagnostic tool that can differentiate between MCI, AD, and other diseases in OCT scans will require larger, comprehensive, and open-access datasets.

\section*{CRediT authorship contribution statement} 

\textbf{Yasemin Turkan} Concept, methodology, invest, data curation, writing org.draft \textbf{F. Boray Tek}  concept, method,  writing review editing, supervision

\section*{Acknowledgements} 
This survey was done as a part of the UK Biobank approved project (Id: 82266) titled ``Deep Learning-Based Analysis of Retinal OCT Scans for Detection of Alzheimer's Disease''.


\bibliographystyle{elsarticle-harv} 
\bibliography{survey-refs}

\begin{thebibliography}{122}
\expandafter\ifx\csname natexlab\endcsname\relax\def\natexlab#1{#1}\fi
\providecommand{\url}[1]{\texttt{#1}}
\providecommand{\href}[2]{#2}
\providecommand{\path}[1]{#1}
\providecommand{\DOIprefix}{doi:}
\providecommand{\ArXivprefix}{arXiv:}
\providecommand{\URLprefix}{URL: }
\providecommand{\Pubmedprefix}{pmid:}
\providecommand{\doi}[1]{\href{http://dx.doi.org/#1}{\path{#1}}}
\providecommand{\Pubmed}[1]{\href{pmid:#1}{\path{#1}}}
\providecommand{\bibinfo}[2]{#2}
\ifx\xfnm\relax \def\xfnm[#1]{\unskip,\space#1}\fi
\bibitem[{ADNI(2022)}]{ADNI}
\bibinfo{author}{ADNI}, \bibinfo{year}{2022}.
\newblock \bibinfo{title}{Alzheimer's disease neuroimaging initiative (adni)}.
\newblock \URLprefix \url{http://adni.loni.usc.edu/}. \bibinfo{note}{[Online;
  accessed 2022-08-13]}.
\bibitem[{Al-Nuaimi et~al.(2021)Al-Nuaimi, Blūma, Al-Juboori, Eke, Jammeh, Sun
  and Ifeachor}]{Al-Nuaimi2021}
\bibinfo{author}{Al-Nuaimi, A.H.}, \bibinfo{author}{Blūma, M.},
  \bibinfo{author}{Al-Juboori, S.S.}, \bibinfo{author}{Eke, C.S.},
  \bibinfo{author}{Jammeh, E.}, \bibinfo{author}{Sun, L.},
  \bibinfo{author}{Ifeachor, E.}, \bibinfo{year}{2021}.
\newblock \bibinfo{title}{{Robust EEG-based biomarkers to detect alzheimer's
  disease}}.
\newblock \bibinfo{journal}{Brain Sciences} \bibinfo{volume}{11}.
\newblock \DOIprefix\doi{10.3390/brainsci11081026}.
\bibitem[{Alber et~al.(2020)Alber, Arthur, Sinoff, {Cabrera Debuc}, Chew,
  Douquette, Hatch, Hudson, Kashani, Lee, Montaquila, Mozdbar, Cunha, Tayyari,
  Stavern and Snyder}]{Alber2020}
\bibinfo{author}{Alber, J.}, \bibinfo{author}{Arthur, E.},
  \bibinfo{author}{Sinoff, S.}, \bibinfo{author}{{Cabrera Debuc}, D.},
  \bibinfo{author}{Chew, E.Y.}, \bibinfo{author}{Douquette, L.},
  \bibinfo{author}{Hatch, W.V.}, \bibinfo{author}{Hudson, C.},
  \bibinfo{author}{Kashani, A.}, \bibinfo{author}{Lee, C.S.},
  \bibinfo{author}{Montaquila, S.}, \bibinfo{author}{Mozdbar, S.},
  \bibinfo{author}{Cunha, L.P.}, \bibinfo{author}{Tayyari, F.},
  \bibinfo{author}{Stavern, G.V.}, \bibinfo{author}{Snyder, P.J.},
  \bibinfo{year}{2020}.
\newblock \bibinfo{title}{{A recommended "minimum data set" framework for
  SD-OCT retinal image acquisition and analysis from the Atlas of Retinal
  Imaging in Alzheimer's Study (ARIAS)}}.
\newblock \bibinfo{journal}{Alzheimer's \& Dementia} \URLprefix
  \url{https://doi.org/10.1002/dad2.12119}, \DOIprefix\doi{10.1002/dad2.12119}.
\bibitem[{Alber et~al.(2020b)Alber, Arthur, Thompson, Goldfarb, Fernandez,
  Salloway, Sinoff and Snyder}]{Alber2020b}
\bibinfo{author}{Alber, J.}, \bibinfo{author}{Arthur, E.},
  \bibinfo{author}{Thompson, L.I.}, \bibinfo{author}{Goldfarb, D.},
  \bibinfo{author}{Fernandez, B.M.}, \bibinfo{author}{Salloway, S.P.},
  \bibinfo{author}{Sinoff, S.}, \bibinfo{author}{Snyder, P.J.},
  \bibinfo{year}{2020b}.
\newblock \bibinfo{title}{{The Atlas of Retinal Imaging in Alzheimer's Study
  (ARIAS): Study design and objectives}}.
\newblock \bibinfo{journal}{Alzheimer's \& Dementia} \bibinfo{volume}{16}.
\newblock \DOIprefix\doi{10.1002/ALZ.045325}.
\bibitem[{Armstrong(2011)}]{Armstrong2011}
\bibinfo{author}{Armstrong, R.A.}, \bibinfo{year}{2011}.
\newblock \bibinfo{title}{{The pathogenesis of alzheimer's disease: A
  reevaluation of the "amyloid cascade hypothesis"}}.
\newblock \bibinfo{journal}{International Journal of Alzheimer's Disease}
  \DOIprefix\doi{10.4061/2011/630865}.
\bibitem[{Attiku et~al.(2021)Attiku, He, Nittala and Sadda}]{Attiku2021}
\bibinfo{author}{Attiku, Y.}, \bibinfo{author}{He, Y.},
  \bibinfo{author}{Nittala, M.}, \bibinfo{author}{Sadda, S.},
  \bibinfo{year}{2021}.
\newblock \bibinfo{title}{{Current status and future possibilities of retinal
  imaging in diabetic retinopathy care applicable to low- and medium-income
  countries}}.
\newblock \bibinfo{journal}{Indian Journal of Ophthalmology}
  \bibinfo{volume}{69}, \bibinfo{pages}{2968}.
\bibitem[{Augustin and Atorf(2022)}]{Augustin2022}
\bibinfo{author}{Augustin, A.J.}, \bibinfo{author}{Atorf, J.},
  \bibinfo{year}{2022}.
\newblock \bibinfo{title}{{The Value of Optical Coherence Tomography
  Angiography (OCT-A) in Neurological Diseases}}.
\newblock \bibinfo{journal}{Diagnostics} \bibinfo{volume}{12}.
\newblock \DOIprefix\doi{10.3390/diagnostics12020468}.
\bibitem[{Bayhan et~al.(2015)Bayhan, {Aslan Bayhan}, Celikbilek, Tanik and
  G{\"{u}}rdal}]{Bayhan2015}
\bibinfo{author}{Bayhan, H.A.}, \bibinfo{author}{{Aslan Bayhan}, S.},
  \bibinfo{author}{Celikbilek, A.}, \bibinfo{author}{Tanik, N.},
  \bibinfo{author}{G{\"{u}}rdal, C.}, \bibinfo{year}{2015}.
\newblock \bibinfo{title}{{Evaluation of the chorioretinal thickness changes in
  Alzheimer's disease using spectral-domain optical coherence tomography}}.
\newblock \bibinfo{journal}{Clinical and Experimental Ophthalmology}
  \bibinfo{volume}{43}, \bibinfo{pages}{145--151}.
\newblock \DOIprefix\doi{10.1111/CEO.12386}.
\bibitem[{Bear et~al.(2016)Bear, Connors and Paradiso}]{Bear2016}
\bibinfo{author}{Bear, M.F.}, \bibinfo{author}{Connors, B.W.},
  \bibinfo{author}{Paradiso, M.A.}, \bibinfo{year}{2016}.
\newblock \bibinfo{title}{{Neuroscience , Exploring the Brain}}.
\newblock \bibinfo{edition}{Fourth edition} ed., \bibinfo{publisher}{Wolters
  Kluwer}.
\bibitem[{Bernardes et~al.(2017)Bernardes, Silva, Chiquita, Serranho and
  Ambr{\'{o}}sio}]{Bernardes2017}
\bibinfo{author}{Bernardes, R.}, \bibinfo{author}{Silva, G.},
  \bibinfo{author}{Chiquita, S.}, \bibinfo{author}{Serranho, P.},
  \bibinfo{author}{Ambr{\'{o}}sio, A.F.}, \bibinfo{year}{2017}.
\newblock \bibinfo{title}{{Retinal biomarkers of Alzheimer's disease: Insights
  from transgenic mouse models}}.
\newblock \bibinfo{journal}{Lecture Notes in Computer Science (including
  subseries Lecture Notes in Artificial Intelligence and Lecture Notes in
  Bioinformatics)} \bibinfo{volume}{10317 LNCS}, \bibinfo{pages}{541--550}.
\newblock \DOIprefix\doi{10.1007/978-3-319-59876-5\_60}.
\bibitem[{Bissig et~al.(2020a)Bissig, Zhou, Le and Bernard}]{Bissig}
\bibinfo{author}{Bissig, D.}, \bibinfo{author}{Zhou, C.}, \bibinfo{author}{Le,
  V.}, \bibinfo{author}{Bernard, J.}, \bibinfo{year}{2020}a.
\newblock \bibinfo{title}{{Dryad Data -- A practical approach to functional
  optical coherence tomography shows abnormal retinal responses in Alzheimer's
  disease}}.
\newblock \bibinfo{journal}{Dryad Digital Repository} \URLprefix
  \url{https://datadryad.org/stash/dataset/doi:10.5061/dryad.msbcc2ftc},
  \DOIprefix\doi{https://doi.org/10.5061/dryad.msbcc2ftc}.
\bibitem[{Bissig et~al.(2020b)Bissig, Zhou, Le and Bernard}]{Bissig2020}
\bibinfo{author}{Bissig, D.}, \bibinfo{author}{Zhou, C.G.},
  \bibinfo{author}{Le, V.}, \bibinfo{author}{Bernard, J.T.},
  \bibinfo{year}{2020}b.
\newblock \bibinfo{title}{{Optical coherence tomography reveals light-dependent
  retinal responses in Alzheimer's disease}}.
\newblock \bibinfo{journal}{NeuroImage} \bibinfo{volume}{219}.
\newblock \URLprefix \url{https://doi.org/10.1016/j.neuroimage.2020.117022},
  \DOIprefix\doi{10.1016/J.NEUROIMAGE.2020.117022}.
\bibitem[{Blazes and Lee()}]{Blazes}
\bibinfo{author}{Blazes, M.}, \bibinfo{author}{Lee, C.S.}, .
\newblock \bibinfo{title}{{Understanding the Brain through Aging Eyes}}.
\newblock \bibinfo{journal}{Advances in Geriatric Medicine and Research}
  \DOIprefix\doi{10.20900/agmr20210008}.
\bibitem[{Bourkhime et~al.(2022)Bourkhime, Qarmiche, Omari, Bahra, Tachfouti,
  Fakir and Otmani}]{Bourkhime2022}
\bibinfo{author}{Bourkhime, H.}, \bibinfo{author}{Qarmiche, N.},
  \bibinfo{author}{Omari, M.}, \bibinfo{author}{Bahra, N.},
  \bibinfo{author}{Tachfouti, N.}, \bibinfo{author}{Fakir, S.E.},
  \bibinfo{author}{Otmani, N.}, \bibinfo{year}{2022}.
\newblock \bibinfo{title}{{Machine learning and novel ophthalmologic biomarkers
  for Alzheimer's disease screening: Systematic Review}}.
\newblock \bibinfo{journal}{ITM Web of Conferences} \bibinfo{volume}{43},
  \bibinfo{pages}{01009}.
\newblock \DOIprefix\doi{10.1051/ITMCONF/20224301009}.
\bibitem[{Budson and Kowall(2011)}]{Budson2011}
\bibinfo{author}{Budson, A.E.}, \bibinfo{author}{Kowall, N.W.},
  \bibinfo{year}{2011}.
\newblock \bibinfo{title}{{The handbook of Alzheimer's disease and other
  dementias}}.
\newblock \bibinfo{publisher}{Wiley-Blackwell}.
\bibitem[{Bulut et~al.(2018)Bulut, Kurtuluş, G{\"{o}}zkaya, Erol, Cengiz,
  Akldan and Yaman}]{Bulut2018}
\bibinfo{author}{Bulut, M.}, \bibinfo{author}{Kurtuluş, F.},
  \bibinfo{author}{G{\"{o}}zkaya, O.}, \bibinfo{author}{Erol, M.K.},
  \bibinfo{author}{Cengiz, A.}, \bibinfo{author}{Akldan, M.},
  \bibinfo{author}{Yaman, A.}, \bibinfo{year}{2018}.
\newblock \bibinfo{title}{{Evaluation of optical coherence tomography
  angiographic findings in Alzheimer's type dementia}}.
\newblock \bibinfo{journal}{British Journal of Ophthalmology}
  \bibinfo{volume}{102}, \bibinfo{pages}{233--237}.
\newblock \DOIprefix\doi{10.1136/BJOPHTHALMOL-2017-310476}.
\bibitem[{Bulut et~al.(2016)Bulut, Yaman, Erol, Kurtuluş, Toslak, Doǧan,
  {\c{C}}oban and Başar}]{Bulut2016}
\bibinfo{author}{Bulut, M.}, \bibinfo{author}{Yaman, A.},
  \bibinfo{author}{Erol, M.K.}, \bibinfo{author}{Kurtuluş, F.},
  \bibinfo{author}{Toslak, D.}, \bibinfo{author}{Doǧan, B.},
  \bibinfo{author}{{\c{C}}oban, D.T.}, \bibinfo{author}{Başar, E.K.},
  \bibinfo{year}{2016}.
\newblock \bibinfo{title}{{Choroidal Thickness in Patients with Mild Cognitive
  Impairment and Alzheimer's Type Dementia}}.
\newblock \bibinfo{journal}{Journal of Ophthalmology} \bibinfo{volume}{2016}.
\newblock \DOIprefix\doi{10.1155/2016/7291257}.
\bibitem[{Carelli et~al.(2017)Carelli, Morgia, Ross-Cisneros and
  Sadun}]{Carelli2017}
\bibinfo{author}{Carelli, V.}, \bibinfo{author}{Morgia, C.L.},
  \bibinfo{author}{Ross-Cisneros, F.N.}, \bibinfo{author}{Sadun, A.A.},
  \bibinfo{year}{2017}.
\newblock \bibinfo{title}{{Optic neuropathies: the tip of the neurodegeneration
  iceberg}}.
\newblock \bibinfo{journal}{Human Molecular Genetics} \URLprefix
  \url{https://academic.oup.com/hmg/article/26/R2/R139/4036433},
  \DOIprefix\doi{10.1093/hmg/ddx273}.
\bibitem[{Chalkias et~al.(2021)Chalkias, Tegos, Topouzis and
  Tsolaki}]{Chalkias2021a}
\bibinfo{author}{Chalkias, I.N.}, \bibinfo{author}{Tegos, T.},
  \bibinfo{author}{Topouzis, F.}, \bibinfo{author}{Tsolaki, M.},
  \bibinfo{year}{2021}.
\newblock \bibinfo{title}{{Ocular biomarkers and their role in the early
  diagnosis of neurocognitive disorders}}.
\newblock \bibinfo{journal}{European Journal of Ophthalmology}
  \bibinfo{volume}{31}, \bibinfo{pages}{2808--2817}.
\newblock \DOIprefix\doi{10.1177/11206721211016311}.
\bibitem[{Choi et~al.(2021)Choi, Choi, Roh, Eun, Kim, Shin, Kang, Chung, Lee,
  Lee, Kang, Cho and Kim}]{Choi2021}
\bibinfo{author}{Choi, K.J.}, \bibinfo{author}{Choi, J.E.},
  \bibinfo{author}{Roh, H.C.}, \bibinfo{author}{Eun, J.S.},
  \bibinfo{author}{Kim, J.M.}, \bibinfo{author}{Shin, Y.K.},
  \bibinfo{author}{Kang, M.C.}, \bibinfo{author}{Chung, J.K.},
  \bibinfo{author}{Lee, C.}, \bibinfo{author}{Lee, D.}, \bibinfo{author}{Kang,
  S.W.}, \bibinfo{author}{Cho, B.H.}, \bibinfo{author}{Kim, S.J.},
  \bibinfo{year}{2021}.
\newblock \bibinfo{title}{{Deep learning models for screening of high myopia
  using optical coherence tomography}}.
\newblock \bibinfo{journal}{Scientific Reports} \bibinfo{volume}{11},
  \bibinfo{pages}{21663}.
\newblock \URLprefix \url{https://www.nature.com/articles/s41598-021-00622-x},
  \DOIprefix\doi{10.1038/S41598-021-00622-X}.
\bibitem[{Chua et~al.(2019)Chua, Thomas, Allen, Lotery, Desai, Patel, Muthy,
  Sudlow, Peto, Khaw, Foster, Zheng, Aslam, Barman, Barrett, Bishop, Blows,
  Bunce, Carare, Chakravarthy, Chan, Crabb, Cumberland, Day, Dhillon, Dick,
  Egan, Ennis, Fruttiger, Gallacher, Garway-Heath, Gibson, Gore, Guggenheim,
  Hammond, Hardcastle, Harding, Hogg, Hysi, Keane, Khawaja, Lascaratos,
  MacGillivray, Mackie, Martin, McGaughey, McGuinness, McKay, McKibbin, Mitry,
  Moore, Morgan, O'Sullivan, Owen, Paterson, Petzold, Rahi, Rudnikca, Self,
  Sivaprasad, Steel, Stratton, Strouthidis, Trucco, Tufail, Vitart, Vernon,
  Viswanathan, Williams, Williams, Woodside, Yates and Yip}]{Chua2019}
\bibinfo{author}{Chua, S.Y.L.}, \bibinfo{author}{Thomas, D.},
  \bibinfo{author}{Allen, N.}, \bibinfo{author}{Lotery, A.},
  \bibinfo{author}{Desai, P.}, \bibinfo{author}{Patel, P.},
  \bibinfo{author}{Muthy, Z.}, \bibinfo{author}{Sudlow, C.},
  \bibinfo{author}{Peto, T.}, \bibinfo{author}{Khaw, P.T.},
  \bibinfo{author}{Foster, P.J.}, \bibinfo{author}{Zheng, Y.},
  \bibinfo{author}{Aslam, T.}, \bibinfo{author}{Barman, S.A.},
  \bibinfo{author}{Barrett, J.H.}, \bibinfo{author}{Bishop, P.},
  \bibinfo{author}{Blows, P.}, \bibinfo{author}{Bunce, C.},
  \bibinfo{author}{Carare, R.O.}, \bibinfo{author}{Chakravarthy, U.},
  \bibinfo{author}{Chan, M.}, \bibinfo{author}{Crabb, D.P.},
  \bibinfo{author}{Cumberland, P.M.}, \bibinfo{author}{Day, A.},
  \bibinfo{author}{Dhillon, B.}, \bibinfo{author}{Dick, A.D.},
  \bibinfo{author}{Egan, C.}, \bibinfo{author}{Ennis, S.},
  \bibinfo{author}{Fruttiger, M.}, \bibinfo{author}{Gallacher, J.E.},
  \bibinfo{author}{Garway-Heath, D.F.}, \bibinfo{author}{Gibson, J.},
  \bibinfo{author}{Gore, D.}, \bibinfo{author}{Guggenheim, J.A.},
  \bibinfo{author}{Hammond, C.J.}, \bibinfo{author}{Hardcastle, A.},
  \bibinfo{author}{Harding, S.P.}, \bibinfo{author}{Hogg, R.E.},
  \bibinfo{author}{Hysi, P.}, \bibinfo{author}{Keane, P.A.},
  \bibinfo{author}{Khawaja, A.P.}, \bibinfo{author}{Lascaratos, G.},
  \bibinfo{author}{MacGillivray, T.}, \bibinfo{author}{Mackie, S.},
  \bibinfo{author}{Martin, K.}, \bibinfo{author}{McGaughey, M.},
  \bibinfo{author}{McGuinness, B.}, \bibinfo{author}{McKay, G.J.},
  \bibinfo{author}{McKibbin, M.}, \bibinfo{author}{Mitry, D.},
  \bibinfo{author}{Moore, T.}, \bibinfo{author}{Morgan, J.E.},
  \bibinfo{author}{O'Sullivan, E.}, \bibinfo{author}{Owen, C.G.},
  \bibinfo{author}{Paterson, E.}, \bibinfo{author}{Petzold, A.},
  \bibinfo{author}{Rahi, J.S.}, \bibinfo{author}{Rudnikca, A.R.},
  \bibinfo{author}{Self, J.}, \bibinfo{author}{Sivaprasad, S.},
  \bibinfo{author}{Steel, D.}, \bibinfo{author}{Stratton, I.},
  \bibinfo{author}{Strouthidis, N.}, \bibinfo{author}{Trucco, E.},
  \bibinfo{author}{Tufail, A.}, \bibinfo{author}{Vitart, V.},
  \bibinfo{author}{Vernon, S.A.}, \bibinfo{author}{Viswanathan, A.C.},
  \bibinfo{author}{Williams, C.}, \bibinfo{author}{Williams, K.},
  \bibinfo{author}{Woodside, J.V.}, \bibinfo{author}{Yates, M.M.},
  \bibinfo{author}{Yip, J.}, \bibinfo{year}{2019}.
\newblock \bibinfo{title}{{Cohort profile: Design and methods in the eye and
  vision consortium of UK Biobank}}.
\newblock \bibinfo{journal}{BMJ Open} \bibinfo{volume}{9},
  \bibinfo{pages}{1--13}.
\newblock \DOIprefix\doi{10.1136/bmjopen-2018-025077}.
\bibitem[{Chueh et~al.(2021)Chueh, Hsieh, Chen, Ma and Huang}]{Chueh2021}
\bibinfo{author}{Chueh, K.M.}, \bibinfo{author}{Hsieh, Y.T.},
  \bibinfo{author}{Chen, H.H.}, \bibinfo{author}{Ma, I.H.},
  \bibinfo{author}{Huang, S.L.}, \bibinfo{year}{2021}.
\newblock \bibinfo{title}{{Identification of Sex and Age from Macular Optical
  Coherence Tomography and Feature Analysis Using Deep Learning}}.
\newblock \bibinfo{journal}{American Journal of Ophthalmology}
  \DOIprefix\doi{10.1016/J.AJO.2021.09.015}.
\bibitem[{ClinicalTrials(2022)}]{ClinicalTrials}
\bibinfo{author}{ClinicalTrials}, \bibinfo{year}{2022}.
\newblock \bibinfo{title}{Clinicaltrials: a website and online database of
  clinical research studies and information about their results}.
\newblock \URLprefix \url{https://www.clinicaltrials.gov/}.
  \bibinfo{note}{[Online; accessed 2022-08-13]}.
\bibitem[{Cootes et~al.(1995)Cootes, Taylor, Cooper and Graham}]{Cootes1995}
\bibinfo{author}{Cootes, T.}, \bibinfo{author}{Taylor, C.},
  \bibinfo{author}{Cooper, D.}, \bibinfo{author}{Graham, J.},
  \bibinfo{year}{1995}.
\newblock \bibinfo{title}{{Cootes\_1995.pdf}}.
\bibitem[{Corbin and Lesage(2022)}]{Corbin2022}
\bibinfo{author}{Corbin, D.}, \bibinfo{author}{Lesage, F.},
  \bibinfo{year}{2022}.
\newblock \bibinfo{title}{{Assessment of the predictive potential of cognitive
  scores from retinal images and retinal fundus metadata via deep learning
  using the CLSA database}}.
\newblock \bibinfo{journal}{Scientific Reports} \bibinfo{volume}{12},
  \bibinfo{pages}{1--12}.
\newblock \URLprefix \url{https://doi.org/10.1038/s41598-022-09719-3},
  \DOIprefix\doi{10.1038/s41598-022-09719-3}.
\bibitem[{Cunha et~al.(2022)Cunha, Pires, Cruzeiro, Almeida, Martins, Martins,
  Shigaeff and Vale}]{Cunha2022}
\bibinfo{author}{Cunha, L.P.}, \bibinfo{author}{Pires, L.A.},
  \bibinfo{author}{Cruzeiro, M.M.}, \bibinfo{author}{Almeida, A.L.M.},
  \bibinfo{author}{Martins, L.C.}, \bibinfo{author}{Martins, P.N.},
  \bibinfo{author}{Shigaeff, N.}, \bibinfo{author}{Vale, T.C.},
  \bibinfo{year}{2022}.
\newblock \bibinfo{title}{{Optical coherence tomography in neurodegenerative
  disorders}}.
\newblock \bibinfo{journal}{Arquivos de Neuro-Psiquiatria}
  \bibinfo{volume}{80}, \bibinfo{pages}{180--191}.
\newblock \DOIprefix\doi{10.1590/0004-282X-ANP-2021-0134}.
\bibitem[{Danesh et~al.(2021)Danesh, Maghooli, Dehghani and
  Kafieh}]{Danesh2021}
\bibinfo{author}{Danesh, H.}, \bibinfo{author}{Maghooli, K.},
  \bibinfo{author}{Dehghani, A.}, \bibinfo{author}{Kafieh, R.},
  \bibinfo{year}{2021}.
\newblock \bibinfo{title}{{Synthetic OCT data in challenging conditions:
  three-dimensional OCT and presence of abnormalities}}.
\newblock \bibinfo{journal}{Medical \& Biological Engineering \& Computing}
  \URLprefix \url{https://link.springer.com/10.1007/s11517-021-02469-w},
  \DOIprefix\doi{10.1007/s11517-021-02469-w}.
\bibitem[{Dening and Sandilyan(2015)}]{Dening2015}
\bibinfo{author}{Dening, T.}, \bibinfo{author}{Sandilyan, M.B.},
  \bibinfo{year}{2015}.
\newblock \bibinfo{title}{{Dementia: definitions and types}}.
\newblock \bibinfo{journal}{Nursing standard (Royal College of Nursing (Great
  Britain) : 1987)} \bibinfo{volume}{29}, \bibinfo{pages}{37--42}.
\newblock \DOIprefix\doi{10.7748/ns.29.37.37.e9405}.
\bibitem[{Eid et~al.(2022)Eid, Arnould, Gabrielle, Aho, Farnier,
  Creuzot-garcher and Cottin}]{Eid2022}
\bibinfo{author}{Eid, P.}, \bibinfo{author}{Arnould, L.},
  \bibinfo{author}{Gabrielle, P.h.}, \bibinfo{author}{Aho, L.S.},
  \bibinfo{author}{Farnier, M.}, \bibinfo{author}{Creuzot-garcher, C.},
  \bibinfo{author}{Cottin, Y.}, \bibinfo{year}{2022}.
\newblock \bibinfo{title}{{Retinal Microvascular Changes in Familial
  Hypercholesterolemia : Analysis with Swept-Source Optical Coherence
  Tomography Angiography}}.
\newblock \bibinfo{journal}{Journal of Personalized Medicine}
  \DOIprefix\doi{2075-4426/12/6/871}.
\bibitem[{Ellendt et~al.(2017)Ellendt, Vo$\beta$, Kohn, Wagels, Goerlich,
  Drexler, Schneider and Habel}]{Ellendt2017}
\bibinfo{author}{Ellendt, S.}, \bibinfo{author}{Vo$\beta$, B.},
  \bibinfo{author}{Kohn, N.}, \bibinfo{author}{Wagels, L.},
  \bibinfo{author}{Goerlich, K.S.}, \bibinfo{author}{Drexler, E.},
  \bibinfo{author}{Schneider, F.}, \bibinfo{author}{Habel, U.},
  \bibinfo{year}{2017}.
\newblock \bibinfo{title}{{Predicting Stability of Mild Cognitive Impairment
  (MCI): Findings of a Community Based Sample}}.
\newblock \bibinfo{journal}{Current Alzheimer research} \bibinfo{volume}{14},
  \bibinfo{pages}{608--619}.
\newblock \URLprefix \url{https://pubmed.ncbi.nlm.nih.gov/27978792/},
  \DOIprefix\doi{10.2174/1567205014666161213120807}.
\bibitem[{Fan et~al.(2021)Fan, Zhang, Zhu, Ma and Zhu}]{Fan2021}
\bibinfo{author}{Fan, F.}, \bibinfo{author}{Zhang, J.}, \bibinfo{author}{Zhu,
  L.}, \bibinfo{author}{Ma, Z.}, \bibinfo{author}{Zhu, J.},
  \bibinfo{year}{2021}.
\newblock \bibinfo{title}{{Improving cerebral microvascular image quality of
  optical coherence tomography angiography with deep learning-based
  segmentation}}.
\newblock \bibinfo{journal}{Journal of Biophotonics} \bibinfo{volume}{14}.
\newblock \DOIprefix\doi{10.1002/JBIO.202100171}.
\bibitem[{Fereshetian et~al.(2021)Fereshetian, Agranat, Siegel, Ness, Stein and
  Subramanian}]{Fereshetian2021}
\bibinfo{author}{Fereshetian, S.}, \bibinfo{author}{Agranat, J.S.},
  \bibinfo{author}{Siegel, N.}, \bibinfo{author}{Ness, S.},
  \bibinfo{author}{Stein, T.D.}, \bibinfo{author}{Subramanian, M.L.},
  \bibinfo{year}{2021}.
\newblock \bibinfo{title}{{Protein and Imaging Biomarkers in the Eye for Early
  Detection of Alzheimer's Disease}}.
\newblock \bibinfo{journal}{Journal of Alzheimer's Disease Reports}
  \bibinfo{volume}{5}, \bibinfo{pages}{375--387}.
\newblock \DOIprefix\doi{10.3233/ADR-210283}.
\bibitem[{Ferrari et~al.(2017)Ferrari, Huang, Magnani, Ambrosi, Comi and
  Leocani}]{Ferrari2017}
\bibinfo{author}{Ferrari, L.}, \bibinfo{author}{Huang, S.C.},
  \bibinfo{author}{Magnani, G.}, \bibinfo{author}{Ambrosi, A.},
  \bibinfo{author}{Comi, G.}, \bibinfo{author}{Leocani, L.},
  \bibinfo{year}{2017}.
\newblock \bibinfo{title}{{Optical Coherence Tomography Reveals Retinal
  Neuroaxonal Thinning in Frontotemporal Dementia as in Alzheimer's Disease}}.
\newblock \bibinfo{journal}{Journal of Alzheimer's Disease}
  \bibinfo{volume}{56}, \bibinfo{pages}{1101--1107}.
\newblock \DOIprefix\doi{10.3233/JAD-160886}.
\bibitem[{Ferreira et~al.(2022)Ferreira, Serranho, Guimar{\~{a}}es, Trindade,
  Martins, Moreira, Ambr{\'{o}}sio and Branco}]{Ferreira2022}
\bibinfo{author}{Ferreira, H.}, \bibinfo{author}{Serranho, P.},
  \bibinfo{author}{Guimar{\~{a}}es, P.}, \bibinfo{author}{Trindade, R.},
  \bibinfo{author}{Martins, J.}, \bibinfo{author}{Moreira, P.I.},
  \bibinfo{author}{Ambr{\'{o}}sio, A.F.}, \bibinfo{author}{Branco, M.C.},
  \bibinfo{year}{2022}.
\newblock \bibinfo{title}{{Stage ‑ independent biomarkers for Alzheimer ' s
  disease from the living retina : an animal study}}.
\newblock \bibinfo{journal}{Scientific Reports} ,
  \bibinfo{pages}{1--7}\URLprefix
  \url{https://doi.org/10.1038/s41598-022-18113-y},
  \DOIprefix\doi{10.1038/s41598-022-18113-y}.
\bibitem[{Folstein et~al.(1975)Folstein, Folstein and McHugh}]{Folstein1975}
\bibinfo{author}{Folstein, M.F.}, \bibinfo{author}{Folstein, S.E.},
  \bibinfo{author}{McHugh, P.R.}, \bibinfo{year}{1975}.
\newblock \bibinfo{title}{{“Mini-mental state”: A practical method for
  grading the cognitive state of patients for the clinician}}.
\newblock \bibinfo{journal}{Journal of Psychiatric Research}
  \bibinfo{volume}{12}, \bibinfo{pages}{189--198}.
\newblock \DOIprefix\doi{10.1016/0022-3956(75)90026-6}.
\bibitem[{Gardner et~al.(2020)Gardner, Baruah, Vargas, Motamedi, Milner and
  Rylander}]{Gardner2020}
\bibinfo{author}{Gardner, M.R.}, \bibinfo{author}{Baruah, V.},
  \bibinfo{author}{Vargas, G.}, \bibinfo{author}{Motamedi, M.},
  \bibinfo{author}{Milner, T.E.}, \bibinfo{author}{Rylander, H.G.},
  \bibinfo{year}{2020}.
\newblock \bibinfo{title}{{Scattering angle resolved optical coherence
  tomography detects early changes in 3xTg Alzheimer's disease mouse model}}.
\newblock \bibinfo{journal}{Translational Vision Science and Technology}
  \bibinfo{volume}{9}, \bibinfo{pages}{1--14}.
\newblock \DOIprefix\doi{10.1167/TVST.9.5.18}.
\bibitem[{Garrondo(2008)}]{Garrondo}
\bibinfo{author}{Garrondo}, \bibinfo{year}{2008}.
\newblock \bibinfo{title}{{File:Brain-ALZH.png - Wikimedia Commons}}.
\newblock \URLprefix
  \url{https://commons.wikimedia.org/wiki/File:Brain-ALZH.png}.
  \bibinfo{note}{[Online; accessed 2022-08-13]}.
\bibitem[{Gaugler et~al.(2022)Gaugler, James, Johnson, Reimer, Solis, Weuve,
  Buckley and Hohman}]{Gaugler2022}
\bibinfo{author}{Gaugler, J.}, \bibinfo{author}{James, B.},
  \bibinfo{author}{Johnson, T.}, \bibinfo{author}{Reimer, J.},
  \bibinfo{author}{Solis, M.}, \bibinfo{author}{Weuve, J.},
  \bibinfo{author}{Buckley, R.F.}, \bibinfo{author}{Hohman, T.J.},
  \bibinfo{year}{2022}.
\newblock \bibinfo{title}{2022 alzheimer's disease facts and figures.}
\newblock \bibinfo{journal}{Alzheimer's \& dementia : the journal of the
  Alzheimer's Association} \bibinfo{volume}{18}, \bibinfo{pages}{700--789}.
\newblock \URLprefix \url{http://www.ncbi.nlm.nih.gov/pubmed/35289055},
  \DOIprefix\doi{10.1002/alz.12638}.
\bibitem[{Ge et~al.(2021)Ge, Xu, Ou, Qu, Ma, Huang, Shen, Chen, Tan, Zhao and
  Yu}]{Ge2021}
\bibinfo{author}{Ge, Y.J.}, \bibinfo{author}{Xu, W.}, \bibinfo{author}{Ou,
  Y.N.}, \bibinfo{author}{Qu, Y.}, \bibinfo{author}{Ma, Y.H.},
  \bibinfo{author}{Huang, Y.Y.}, \bibinfo{author}{Shen, X.N.},
  \bibinfo{author}{Chen, S.D.}, \bibinfo{author}{Tan, L.},
  \bibinfo{author}{Zhao, Q.H.}, \bibinfo{author}{Yu, J.T.},
  \bibinfo{year}{2021}.
\newblock \bibinfo{title}{{Retinal biomarkers in Alzheimer's disease and mild
  cognitive impairment: A systematic review and meta-analysis}}.
\newblock \bibinfo{journal}{Ageing Research Reviews} \bibinfo{volume}{69}.
\newblock \DOIprefix\doi{10.1016/j.arr.2021.101361}.
\bibitem[{Gharbiya et~al.(2014)Gharbiya, Trebbastoni, Parisi, Manganiello,
  Cruciani, D'Antonio, {De Vico}, Imbriano, Campanelli and {De
  Lena}}]{Gharbiya2014}
\bibinfo{author}{Gharbiya, M.}, \bibinfo{author}{Trebbastoni, A.},
  \bibinfo{author}{Parisi, F.}, \bibinfo{author}{Manganiello, S.},
  \bibinfo{author}{Cruciani, F.}, \bibinfo{author}{D'Antonio, F.},
  \bibinfo{author}{{De Vico}, U.}, \bibinfo{author}{Imbriano, L.},
  \bibinfo{author}{Campanelli, A.}, \bibinfo{author}{{De Lena}, C.},
  \bibinfo{year}{2014}.
\newblock \bibinfo{title}{{Choroidal thinning as a new finding in Alzheimer's
  Disease: Evidence from enhanced depth imaging spectral domain optical
  coherence tomography}}.
\newblock \bibinfo{journal}{Journal of Alzheimer's Disease}
  \bibinfo{volume}{40}, \bibinfo{pages}{907--917}.
\newblock \DOIprefix\doi{10.3233/JAD-132039}.
\bibitem[{Green(2021)}]{Green}
\bibinfo{author}{Green, K.}, \bibinfo{year}{2021}.
\newblock \bibinfo{title}{{File:Color OCTA 2.png - EyeWiki}}.
\newblock \URLprefix \url{https://eyewiki.aao.org/File:Color\_OCTA\_2.png}.
  \bibinfo{note}{[Online; accessed 2022-08-13]}.
\bibitem[{Guimar{\~{a}}es et~al.(2014)Guimar{\~{a}}es, Rodrigues, Lobo, Leal,
  Figueira, Serranho and Bernardes}]{Guimaraes2014}
\bibinfo{author}{Guimar{\~{a}}es, P.}, \bibinfo{author}{Rodrigues, P.},
  \bibinfo{author}{Lobo, C.}, \bibinfo{author}{Leal, S.},
  \bibinfo{author}{Figueira, J.}, \bibinfo{author}{Serranho, P.},
  \bibinfo{author}{Bernardes, R.}, \bibinfo{year}{2014}.
\newblock \bibinfo{title}{{Ocular fundus reference images from optical
  coherence tomography}}.
\newblock \bibinfo{journal}{Computerized Medical Imaging and Graphics}
  \bibinfo{volume}{38}, \bibinfo{pages}{381--389}.
\newblock \DOIprefix\doi{10.1016/j.compmedimag.2014.02.003}.
\bibitem[{den Haan et~al.(2018)den Haan, Morrema, Verbraak, de~Boer, Scheltens,
  Rozemuller, Bergen, Bouwman and Hoozemans}]{DenHaan2018}
\bibinfo{author}{den Haan, J.}, \bibinfo{author}{Morrema, T.H.},
  \bibinfo{author}{Verbraak, F.D.}, \bibinfo{author}{de~Boer, J.F.},
  \bibinfo{author}{Scheltens, P.}, \bibinfo{author}{Rozemuller, A.J.},
  \bibinfo{author}{Bergen, A.A.}, \bibinfo{author}{Bouwman, F.H.},
  \bibinfo{author}{Hoozemans, J.J.}, \bibinfo{year}{2018}.
\newblock \bibinfo{title}{{Amyloid-beta and phosphorylated tau in post-mortem
  Alzheimer's disease retinas}}.
\newblock \bibinfo{journal}{Acta neuropathologica communications}
  \bibinfo{volume}{6}, \bibinfo{pages}{147}.
\newblock \DOIprefix\doi{10.1186/s40478-018-0650-x}.
\bibitem[{{Habib Havoutis}(2017)}]{HabibHavoutis2017}
\bibinfo{author}{{Habib Havoutis}, G.}, \bibinfo{year}{2017}.
\newblock \bibinfo{title}{{NSUWorks Analysis of Diagnostic, Preventive, and
  Disease-Modifying Therapeutic Measures of Alzheimer' s Disease}}.
\newblock \bibinfo{type}{Technical Report}. Capstone Nova Southeastern
  University.
\newblock \URLprefix \url{https://nsuworks.nova.edu/cnso\_stucap/334}.
\bibitem[{Hadoux et~al.(2019)Hadoux, Hui, Lim, Masters, P{\'{e}}bay, Chevalier,
  Ha, Loi, Fowler, Rowe, Villemagne, Taylor, Fluke, Soucy, Lesage, Sylvestre,
  Rosa-Neto, Mathotaarachchi, Gauthier, Nasreddine, Arbour, Rh{\'{e}}aume,
  Beaulieu, Dirani, Nguyen, Bui, Williamson, Crowston and van
  Wijngaarden}]{Hadoux2019}
\bibinfo{author}{Hadoux, X.}, \bibinfo{author}{Hui, F.}, \bibinfo{author}{Lim,
  J.K.}, \bibinfo{author}{Masters, C.L.}, \bibinfo{author}{P{\'{e}}bay, A.},
  \bibinfo{author}{Chevalier, S.}, \bibinfo{author}{Ha, J.},
  \bibinfo{author}{Loi, S.}, \bibinfo{author}{Fowler, C.J.},
  \bibinfo{author}{Rowe, C.}, \bibinfo{author}{Villemagne, V.L.},
  \bibinfo{author}{Taylor, E.N.}, \bibinfo{author}{Fluke, C.},
  \bibinfo{author}{Soucy, J.P.}, \bibinfo{author}{Lesage, F.},
  \bibinfo{author}{Sylvestre, J.P.}, \bibinfo{author}{Rosa-Neto, P.},
  \bibinfo{author}{Mathotaarachchi, S.}, \bibinfo{author}{Gauthier, S.},
  \bibinfo{author}{Nasreddine, Z.S.}, \bibinfo{author}{Arbour, J.D.},
  \bibinfo{author}{Rh{\'{e}}aume, M.A.}, \bibinfo{author}{Beaulieu, S.},
  \bibinfo{author}{Dirani, M.}, \bibinfo{author}{Nguyen, C.T.},
  \bibinfo{author}{Bui, B.V.}, \bibinfo{author}{Williamson, R.},
  \bibinfo{author}{Crowston, J.G.}, \bibinfo{author}{van Wijngaarden, P.},
  \bibinfo{year}{2019}.
\newblock \bibinfo{title}{{Non-invasive in vivo hyperspectral imaging of the
  retina for potential biomarker use in Alzheimer's disease}}.
\newblock \bibinfo{journal}{Nature Communications} \bibinfo{volume}{10}.
\newblock \DOIprefix\doi{10.1038/S41467-019-12242-1}.
\bibitem[{Harzing(2010)}]{Harzing2010}
\bibinfo{author}{Harzing, A.W.}, \bibinfo{year}{2010}.
\newblock \bibinfo{title}{{The Publish or perish book : your guide to effective
  and responsible citation analysis}}.
\newblock \bibinfo{publisher}{Tarma Software Research Pty Ltd}.
\bibitem[{Heidelberg(2010)}]{Heidelberg}
\bibinfo{author}{Heidelberg}, \bibinfo{year}{2010}.
\newblock \bibinfo{title}{{File:Oct SpectralisLayers reduced.jpg - EyeWiki}}.
\newblock \URLprefix
  \url{https://eyewiki.aao.org/File:Oct\_SpectralisLayers\_reduced.jpg}.
  \bibinfo{note}{[Online; accessed 2022-08-13]}.
\bibitem[{Huang et~al.(2021)Huang, Xia, Lu, Liu, Chen, Zhou, Fang and
  Zhang}]{Huang2021}
\bibinfo{author}{Huang, Y.}, \bibinfo{author}{Xia, W.}, \bibinfo{author}{Lu,
  Z.}, \bibinfo{author}{Liu, Y.}, \bibinfo{author}{Chen, H.},
  \bibinfo{author}{Zhou, J.}, \bibinfo{author}{Fang, L.},
  \bibinfo{author}{Zhang, Y.}, \bibinfo{year}{2021}.
\newblock \bibinfo{title}{{Noise-Powered Disentangled Representation for
  Unsupervised Speckle Reduction of Optical Coherence Tomography Images}}.
\newblock \bibinfo{journal}{IEEE Transactions on Medical Imaging}
  \bibinfo{volume}{40}, \bibinfo{pages}{2600--2614}.
\newblock \DOIprefix\doi{10.1109/TMI.2020.3045207},
  \href{http://arxiv.org/abs/2007.03446}{{\tt arXiv:2007.03446}}.
\bibitem[{Hui et~al.(2021)Hui, Zhao, Yu, Liu, Chiu and Wang}]{Hui2021}
\bibinfo{author}{Hui, J.}, \bibinfo{author}{Zhao, Y.}, \bibinfo{author}{Yu,
  S.}, \bibinfo{author}{Liu, J.}, \bibinfo{author}{Chiu, K.},
  \bibinfo{author}{Wang, Y.}, \bibinfo{year}{2021}.
\newblock \bibinfo{title}{{Detection of retinal changes with optical coherence
  tomography angiography in mild cognitive impairment and Alzheimer's disease
  patients: A meta-analysis}}.
\newblock \bibinfo{journal}{PLoS ONE} \bibinfo{volume}{16}.
\newblock \DOIprefix\doi{10.1371/JOURNAL.PONE.0255362}.
\bibitem[{Jonas et~al.(2016)Jonas, Wang, Wei, Zhu, Shao and Xu}]{Jonas2016}
\bibinfo{author}{Jonas, J.B.}, \bibinfo{author}{Wang, Y.X.},
  \bibinfo{author}{Wei, W.B.}, \bibinfo{author}{Zhu, L.P.},
  \bibinfo{author}{Shao, L.}, \bibinfo{author}{Xu, L.}, \bibinfo{year}{2016}.
\newblock \bibinfo{title}{{Cognitive Function and Subfoveal Choroidal
  Thickness: The Beijing Eye Study}}.
\newblock \bibinfo{journal}{Ophthalmology} \bibinfo{volume}{123},
  \bibinfo{pages}{220--222}.
\newblock \DOIprefix\doi{10.1016/j.ophtha.2015.06.020}.
\bibitem[{Juottonen et~al.(1999)Juottonen, Laakso, Partanen and
  Soininen}]{Juottonen1999}
\bibinfo{author}{Juottonen, K.}, \bibinfo{author}{Laakso, M.P.},
  \bibinfo{author}{Partanen, K.}, \bibinfo{author}{Soininen, H.},
  \bibinfo{year}{1999}.
\newblock \bibinfo{title}{{Comparative MR Analysis of the Entorhinal Cortex and
  Hippocampus in Diagnosing Alzheimer Disease}}.
\newblock \bibinfo{journal}{American Journal of Neuroradiology}
  \bibinfo{volume}{20}, \bibinfo{pages}{139--144}.
\bibitem[{Khan et~al.(2021)Khan, Liu, Nath, Korot, Faes, Wagner, Keane, Sebire,
  Burton and Denniston}]{Khan2021}
\bibinfo{author}{Khan, S.M.}, \bibinfo{author}{Liu, X.}, \bibinfo{author}{Nath,
  S.}, \bibinfo{author}{Korot, E.}, \bibinfo{author}{Faes, L.},
  \bibinfo{author}{Wagner, S.K.}, \bibinfo{author}{Keane, P.A.},
  \bibinfo{author}{Sebire, N.J.}, \bibinfo{author}{Burton, M.J.},
  \bibinfo{author}{Denniston, A.K.}, \bibinfo{year}{2021}.
\newblock \bibinfo{title}{{A global review of publicly available datasets for
  ophthalmological imaging: barriers to access, usability, and
  generalisability}}.
\newblock \DOIprefix\doi{10.1016/S2589-7500(20)30240-5}.
\bibitem[{Kim et~al.(2022)Kim, Han, Park, Bae, Woo and Kim}]{Kim2022}
\bibinfo{author}{Kim, H.M.}, \bibinfo{author}{Han, J.W.},
  \bibinfo{author}{Park, Y.J.}, \bibinfo{author}{Bae, J.B.},
  \bibinfo{author}{Woo, S.J.}, \bibinfo{author}{Kim, K.W.},
  \bibinfo{year}{2022}.
\newblock \bibinfo{title}{{Association Between Retinal Layer Thickness and
  Cognitive Decline in Older Adults}}.
\newblock \bibinfo{journal}{JAMA Ophthalmology} \bibinfo{volume}{140},
  \bibinfo{pages}{683--690}.
\newblock \DOIprefix\doi{10.1001/jamaophthalmol.2022.1563}.
\bibitem[{Latha et~al.(2021)Latha, Sreekanth, Bizu, Suvalakshmi and
  Selvaraj}]{Latha2021}
\bibinfo{author}{Latha, R.S.}, \bibinfo{author}{Sreekanth, G.R.},
  \bibinfo{author}{Bizu, B.}, \bibinfo{author}{Suvalakshmi, K.},
  \bibinfo{author}{Selvaraj, R.E.}, \bibinfo{year}{2021}.
\newblock \bibinfo{title}{{Analysis Of Ophthalmic Disorders For Retinal Images
  Using Deep Learning: A Review}}.
\newblock \bibinfo{journal}{Natural Volatiles \& Essential Oils}
  \bibinfo{volume}{8}, \bibinfo{pages}{656--673}.
\newblock \URLprefix
  \url{https://www.nveo.org/index.php/journal/article/view/434/402}.
\bibitem[{Lecun et~al.(2015)Lecun, Bengio and Hinton}]{Lecun2015}
\bibinfo{author}{Lecun, Y.}, \bibinfo{author}{Bengio, Y.},
  \bibinfo{author}{Hinton, G.}, \bibinfo{year}{2015}.
\newblock \bibinfo{title}{{Deep learning}}.
\newblock \DOIprefix\doi{10.1038/nature14539}.
\bibitem[{Lee et~al.(2017)Lee, Tyring, Deruyter, Wu, Rokem and Lee}]{Lee2017}
\bibinfo{author}{Lee, C.S.}, \bibinfo{author}{Tyring, A.J.},
  \bibinfo{author}{Deruyter, N.P.}, \bibinfo{author}{Wu, Y.},
  \bibinfo{author}{Rokem, A.}, \bibinfo{author}{Lee, A.Y.},
  \bibinfo{year}{2017}.
\newblock \bibinfo{title}{{Deep-learning based, automated segmentation of
  macular edema in optical coherence tomography}}.
\newblock \bibinfo{journal}{Biomed Opt Express.} \bibinfo{volume}{8},
  \bibinfo{pages}{3440--3448}.
\newblock \DOIprefix\doi{10.1364/boe.8.003440}.
\bibitem[{Lee et~al.(2018)Lee, Abraham, Swenor, Sharrett and Ramulu}]{Lee2018}
\bibinfo{author}{Lee, M.J.}, \bibinfo{author}{Abraham, A.G.},
  \bibinfo{author}{Swenor, B.K.}, \bibinfo{author}{Sharrett, A.R.},
  \bibinfo{author}{Ramulu, P.Y.}, \bibinfo{year}{2018}.
\newblock \bibinfo{title}{{Application of optical coherence tomography in the
  detection and classification of cognitive decline}}.
\newblock \bibinfo{journal}{Journal of Current Glaucoma Practice}
  \bibinfo{volume}{12}, \bibinfo{pages}{10--18}.
\newblock \DOIprefix\doi{10.5005/jp-journals-10028-1238}.
\bibitem[{Lemmens et~al.(2020)Lemmens, {Van Craenendonck}, {Van Eijgen}, {De
  Groef}, Bruffaerts, de~Jesus, Charle, Jayapala, Sunaric-M{\'{e}}gevand,
  Standaert, Theunis, {Van Keer}, Vandenbulcke, Moons, Vandenberghe, {De
  Boever} and Stalmans}]{Lemmens2020}
\bibinfo{author}{Lemmens, S.}, \bibinfo{author}{{Van Craenendonck}, T.},
  \bibinfo{author}{{Van Eijgen}, J.}, \bibinfo{author}{{De Groef}, L.},
  \bibinfo{author}{Bruffaerts, R.}, \bibinfo{author}{de~Jesus, D.A.},
  \bibinfo{author}{Charle, W.}, \bibinfo{author}{Jayapala, M.},
  \bibinfo{author}{Sunaric-M{\'{e}}gevand, G.}, \bibinfo{author}{Standaert,
  A.}, \bibinfo{author}{Theunis, J.}, \bibinfo{author}{{Van Keer}, K.},
  \bibinfo{author}{Vandenbulcke, M.}, \bibinfo{author}{Moons, L.},
  \bibinfo{author}{Vandenberghe, R.}, \bibinfo{author}{{De Boever}, P.},
  \bibinfo{author}{Stalmans, I.}, \bibinfo{year}{2020}.
\newblock \bibinfo{title}{{Combination of snapshot hyperspectral retinal
  imaging and optical coherence tomography to identify Alzheimer's disease
  patients}}.
\newblock \bibinfo{journal}{Alzheimer's Research and Therapy}
  \bibinfo{volume}{12}.
\newblock \DOIprefix\doi{10.1186/s13195-020-00715-1}.
\bibitem[{London et~al.(2013)London, Benhar and Schwartz}]{London2013}
\bibinfo{author}{London, A.}, \bibinfo{author}{Benhar, I.},
  \bibinfo{author}{Schwartz, M.}, \bibinfo{year}{2013}.
\newblock \bibinfo{title}{{The retina as a window to the brain - From eye
  research to CNS disorders}}.
\newblock \bibinfo{journal}{Nature Reviews Neurology} \bibinfo{volume}{9},
  \bibinfo{pages}{44--53}.
\newblock \DOIprefix\doi{10.1038/NRNEUROL.2012.227}.
\bibitem[{L{\'{o}}pez-De-Eguileta et~al.(2020)L{\'{o}}pez-De-Eguileta, Lage,
  L{\'{o}}pez-Garc{\'{i}}a, Pozueta, Garc{\'{i}}a-Mart{\'{i}}nez, Kazimierczak,
  Bravo, de~Arcocha-Torres, Banzo, Jimenez-Bonilla, Cerver{\'{o}}, Goikoetxea,
  Rodr{\'{i}}guez-Rodr{\'{i}}guez, S{\'{a}}nchez-Juan and
  Casado}]{Lopez-De-Eguileta2020}
\bibinfo{author}{L{\'{o}}pez-De-Eguileta, A.}, \bibinfo{author}{Lage, C.},
  \bibinfo{author}{L{\'{o}}pez-Garc{\'{i}}a, S.}, \bibinfo{author}{Pozueta,
  A.}, \bibinfo{author}{Garc{\'{i}}a-Mart{\'{i}}nez, M.},
  \bibinfo{author}{Kazimierczak, M.}, \bibinfo{author}{Bravo, M.},
  \bibinfo{author}{de~Arcocha-Torres, M.}, \bibinfo{author}{Banzo, I.},
  \bibinfo{author}{Jimenez-Bonilla, J.}, \bibinfo{author}{Cerver{\'{o}}, A.},
  \bibinfo{author}{Goikoetxea, A.},
  \bibinfo{author}{Rodr{\'{i}}guez-Rodr{\'{i}}guez, E.},
  \bibinfo{author}{S{\'{a}}nchez-Juan, P.}, \bibinfo{author}{Casado, A.},
  \bibinfo{year}{2020}.
\newblock \bibinfo{title}{{Evaluation of choroidal thickness in prodromal
  Alzheimer's disease defined by amyloid PET}}.
\newblock \bibinfo{journal}{PLoS ONE} \bibinfo{volume}{15}.
\newblock \DOIprefix\doi{10.1371/journal.pone.0239484}.
\bibitem[{Ma et~al.(2021)Ma, Hao, Xie, Fu, Zhang, Yang, Wang, Liu, Zheng and
  Zhao}]{Ma2021}
\bibinfo{author}{Ma, Y.}, \bibinfo{author}{Hao, H.}, \bibinfo{author}{Xie, J.},
  \bibinfo{author}{Fu, H.}, \bibinfo{author}{Zhang, J.}, \bibinfo{author}{Yang,
  J.}, \bibinfo{author}{Wang, Z.}, \bibinfo{author}{Liu, J.},
  \bibinfo{author}{Zheng, Y.}, \bibinfo{author}{Zhao, Y.},
  \bibinfo{year}{2021}.
\newblock \bibinfo{title}{{ROSE: A Retinal OCT-Angiography Vessel Segmentation
  Dataset and New Model}}.
\newblock \bibinfo{journal}{IEEE Transactions on Medical Imaging}
  \bibinfo{volume}{40}, \bibinfo{pages}{928--939}.
\newblock \DOIprefix\doi{10.1109/TMI.2020.3042802},
  \href{http://arxiv.org/abs/2007.05201}{{\tt arXiv:2007.05201}}.
\bibitem[{MacGillivray et~al.(2018)MacGillivray, McGrory, Pearson and
  Cameron}]{MacGillivray2018}
\bibinfo{author}{MacGillivray, T.}, \bibinfo{author}{McGrory, S.},
  \bibinfo{author}{Pearson, T.}, \bibinfo{author}{Cameron, J.},
  \bibinfo{year}{2018}.
\newblock \bibinfo{title}{{Retinal imaging in early Alzheimer's disease}}.
\newblock \bibinfo{journal}{Neuromethods} \bibinfo{volume}{137},
  \bibinfo{pages}{199--212}.
\newblock \DOIprefix\doi{10.1007/978-1-4939-7674-4\_14}.
\bibitem[{McKhann et~al.(2011)McKhann, Knopman, Chertkow, Hyman, Jack, Kawas,
  Klunk, Koroshetz, Manly, Mayeux, Mohs, Morris, Rossor, Scheltens, Carrillo,
  Thies, Weintraub and Phelps}]{McKhann2011}
\bibinfo{author}{McKhann, G.M.}, \bibinfo{author}{Knopman, D.S.},
  \bibinfo{author}{Chertkow, H.}, \bibinfo{author}{Hyman, B.T.},
  \bibinfo{author}{Jack, C.R.}, \bibinfo{author}{Kawas, C.H.},
  \bibinfo{author}{Klunk, W.E.}, \bibinfo{author}{Koroshetz, W.J.},
  \bibinfo{author}{Manly, J.J.}, \bibinfo{author}{Mayeux, R.},
  \bibinfo{author}{Mohs, R.C.}, \bibinfo{author}{Morris, J.C.},
  \bibinfo{author}{Rossor, M.N.}, \bibinfo{author}{Scheltens, P.},
  \bibinfo{author}{Carrillo, M.C.}, \bibinfo{author}{Thies, B.},
  \bibinfo{author}{Weintraub, S.}, \bibinfo{author}{Phelps, C.H.},
  \bibinfo{year}{2011}.
\newblock \bibinfo{title}{{The diagnosis of dementia due to Alzheimer's
  disease: Recommendations from the National Institute on Aging-Alzheimer's
  Association workgroups on diagnostic guidelines for Alzheimer's disease}}.
\newblock \bibinfo{journal}{Alzheimer's and Dementia} \bibinfo{volume}{7},
  \bibinfo{pages}{263--269}.
\newblock \DOIprefix\doi{10.1016/j.jalz.2011.03.005}.
\bibitem[{Melin{\v{s}}{\v{c}}ak et~al.(2021)Melin{\v{s}}{\v{c}}ak,
  Radmilovi{\'{c}}, Vatavuk and Lon{\v{c}}ari{\'{c}}}]{Melinscak2021}
\bibinfo{author}{Melin{\v{s}}{\v{c}}ak, M.}, \bibinfo{author}{Radmilovi{\'{c}},
  M.}, \bibinfo{author}{Vatavuk, Z.}, \bibinfo{author}{Lon{\v{c}}ari{\'{c}},
  S.}, \bibinfo{year}{2021}.
\newblock \bibinfo{title}{{Annotated retinal optical coherence tomography
  images (AROI) database for joint retinal layer and fluid segmentation}}.
\newblock \bibinfo{journal}{Automatika} \bibinfo{volume}{62},
  \bibinfo{pages}{375--385}.
\newblock \DOIprefix\doi{10.1080/00051144.2021.1973298}.
\bibitem[{Mirshahi et~al.(2021)Mirshahi, Anvari, Riazi-Esfahani, Sardarinia,
  Naseripour and Falavarjani}]{Mirshahi2021}
\bibinfo{author}{Mirshahi, R.}, \bibinfo{author}{Anvari, P.},
  \bibinfo{author}{Riazi-Esfahani, H.}, \bibinfo{author}{Sardarinia, M.},
  \bibinfo{author}{Naseripour, M.}, \bibinfo{author}{Falavarjani, K.G.},
  \bibinfo{year}{2021}.
\newblock \bibinfo{title}{{Foveal avascular zone segmentation in optical
  coherence tomography angiography images using a deep learning approach}}.
\newblock \bibinfo{journal}{Scientific Reports} \bibinfo{volume}{11},
  \bibinfo{pages}{1--8}.
\newblock \URLprefix \url{https://doi.org/10.1038/s41598-020-80058-x},
  \DOIprefix\doi{10.1038/s41598-020-80058-x}.
\bibitem[{Mirzania et~al.(2021)Mirzania, Thompson and Muir}]{Mirzania2021}
\bibinfo{author}{Mirzania, D.}, \bibinfo{author}{Thompson, A.C.},
  \bibinfo{author}{Muir, K.W.}, \bibinfo{year}{2021}.
\newblock \bibinfo{title}{{Applications of deep learning in detection of
  glaucoma: A systematic review}}.
\newblock \bibinfo{journal}{European Journal of Ophthalmology}
  \bibinfo{volume}{31}, \bibinfo{pages}{1618--1642}.
\newblock \DOIprefix\doi{10.1177/1120672120977346}.
\bibitem[{Montavon et~al.(2017)Montavon, Lapuschkin, Binder, Samek and
  M{\"{u}}ller}]{Montavon2017}
\bibinfo{author}{Montavon, G.}, \bibinfo{author}{Lapuschkin, S.},
  \bibinfo{author}{Binder, A.}, \bibinfo{author}{Samek, W.},
  \bibinfo{author}{M{\"{u}}ller, K.R.}, \bibinfo{year}{2017}.
\newblock \bibinfo{title}{{Explaining nonlinear classification decisions with
  deep Taylor decomposition}}.
\newblock \bibinfo{journal}{Pattern Recognition} \bibinfo{volume}{65},
  \bibinfo{pages}{211--222}.
\newblock \URLprefix \url{http://dx.doi.org/10.1016/j.patcog.2016.11.008},
  \DOIprefix\doi{10.1016/j.patcog.2016.11.008},
  \href{http://arxiv.org/abs/1512.02479}{{\tt arXiv:1512.02479}}.
\bibitem[{Mopuru et~al.(2022)Mopuru, Liu and Arevalo}]{Mopuru2022}
\bibinfo{author}{Mopuru, R.}, \bibinfo{author}{Liu, T.Y.},
  \bibinfo{author}{Arevalo, J.F.}, \bibinfo{year}{2022}.
\newblock \bibinfo{title}{{Choroidal Macrovessel Diagnosed on Multimodal
  Imaging, including Swept-Source Optical Coherence Tomography Angiography}}.
\newblock \bibinfo{journal}{Case Reports in Ophthalmology} ,
  \bibinfo{pages}{215--219}\DOIprefix\doi{10.1159/000521895}.
\bibitem[{Moran et~al.(2022)Moran, Xu, Mehta, Gillies, Karayiannis, Beare, Chen
  and Srikanth}]{Moran2022}
\bibinfo{author}{Moran, C.}, \bibinfo{author}{Xu, Z.Y.},
  \bibinfo{author}{Mehta, H.}, \bibinfo{author}{Gillies, M.},
  \bibinfo{author}{Karayiannis, C.}, \bibinfo{author}{Beare, R.},
  \bibinfo{author}{Chen, C.}, \bibinfo{author}{Srikanth, V.},
  \bibinfo{year}{2022}.
\newblock \bibinfo{title}{{Neuroimaging and cognitive correlates of retinal
  Optical Coherence Tomography (OCT) measures at late middle age in a twin
  sample}}.
\newblock \bibinfo{journal}{Scientific Reports} \bibinfo{volume}{12},
  \bibinfo{pages}{1--9}.
\newblock \URLprefix \url{https://doi.org/10.1038/s41598-022-13662-8},
  \DOIprefix\doi{10.1038/s41598-022-13662-8}.
\bibitem[{Mu et~al.(1999)Mu, Xie, Wen, Weng and Shuyun}]{Mu1999}
\bibinfo{author}{Mu, Q.}, \bibinfo{author}{Xie, J.}, \bibinfo{author}{Wen, Z.},
  \bibinfo{author}{Weng, Y.}, \bibinfo{author}{Shuyun, Z.},
  \bibinfo{year}{1999}.
\newblock \bibinfo{title}{{A Quantitative MR Study of the Hippocampal
  Formation, the Amygdala, and the Temporal Horn of the Lateral Ventricle in
  Healthy Subjects 40 to 90 Years of Age}}.
\newblock \bibinfo{journal}{American Journal of Neuroradiology}
  \bibinfo{volume}{20}, \bibinfo{pages}{207--211}.
\bibitem[{Nguyen et~al.(2020)Nguyen, Ta, Nguyen, Nguyen and Vo}]{Nguyen2020}
\bibinfo{author}{Nguyen, T.T.}, \bibinfo{author}{Ta, Q.T.H.},
  \bibinfo{author}{Nguyen, T.K.O.}, \bibinfo{author}{Nguyen, T.T.D.},
  \bibinfo{author}{Vo, V.G.}, \bibinfo{year}{2020}.
\newblock \bibinfo{title}{{Role of body-fluid biomarkers in Alzheimer's disease
  diagnosis}}.
\newblock \DOIprefix\doi{10.3390/diagnostics10050326}.
\bibitem[{Nunes et~al.(2019)Nunes, Silva, Duque, Janu{\'{a}}rio, Santana,
  Ambr{\'{o}}sio, Castelo-Branco and Bernardes}]{Nunes2019}
\bibinfo{author}{Nunes, A.}, \bibinfo{author}{Silva, G.},
  \bibinfo{author}{Duque, C.}, \bibinfo{author}{Janu{\'{a}}rio, C.},
  \bibinfo{author}{Santana, I.}, \bibinfo{author}{Ambr{\'{o}}sio, A.F.},
  \bibinfo{author}{Castelo-Branco, M.}, \bibinfo{author}{Bernardes, R.},
  \bibinfo{year}{2019}.
\newblock \bibinfo{title}{{Retinal texture biomarkers may help to discriminate
  between Alzheimer's, Parkinson's, and healthy controls}}.
\newblock \bibinfo{journal}{PLoS ONE} \bibinfo{volume}{14}.
\newblock \DOIprefix\doi{10.1371/JOURNAL.PONE.0218826}.
\bibitem[{Obermeyer and Lee(2017)}]{Obermeyer2017}
\bibinfo{author}{Obermeyer, Z.}, \bibinfo{author}{Lee, T.H.},
  \bibinfo{year}{2017}.
\newblock \bibinfo{title}{{Lost in Thought — The Limits of the Human Mind and
  the Future of Medicine}}.
\newblock \bibinfo{journal}{New England Journal of Medicine}
  \bibinfo{volume}{377}, \bibinfo{pages}{1209--1211}.
\newblock \DOIprefix\doi{10.1056/nejmp1705348}.
\bibitem[{O'Bryhim et~al.(2021)O'Bryhim, Lin, {Van Stavern} and
  Apte}]{OBryhim2021}
\bibinfo{author}{O'Bryhim, B.E.}, \bibinfo{author}{Lin, J.B.},
  \bibinfo{author}{{Van Stavern}, G.P.}, \bibinfo{author}{Apte, R.S.},
  \bibinfo{year}{2021}.
\newblock \bibinfo{title}{{OCT Angiography Findings in Preclinical Alzheimer's
  Disease: 3-Year Follow-Up}}.
\newblock \bibinfo{journal}{Ophthalmology} \bibinfo{volume}{128},
  \bibinfo{pages}{1489--1491}.
\newblock \DOIprefix\doi{10.1016/j.ophtha.2021.02.016}.
\bibitem[{Page et~al.(2021)Page, McKenzie, Bossuyt, Boutron, Hoffmann, Mulrow,
  Shamseer, Tetzlaff, Akl, Brennan, Chou, Glanville, Grimshaw,
  Hr{\'{o}}bjartsson, Lalu, Li, Loder, Mayo-Wilson, McDonald, McGuinness,
  Stewart, Thomas, Tricco, Welch, Whiting and Moher}]{Page2021}
\bibinfo{author}{Page, M.J.}, \bibinfo{author}{McKenzie, J.E.},
  \bibinfo{author}{Bossuyt, P.M.}, \bibinfo{author}{Boutron, I.},
  \bibinfo{author}{Hoffmann, T.C.}, \bibinfo{author}{Mulrow, C.D.},
  \bibinfo{author}{Shamseer, L.}, \bibinfo{author}{Tetzlaff, J.M.},
  \bibinfo{author}{Akl, E.A.}, \bibinfo{author}{Brennan, S.E.},
  \bibinfo{author}{Chou, R.}, \bibinfo{author}{Glanville, J.},
  \bibinfo{author}{Grimshaw, J.M.}, \bibinfo{author}{Hr{\'{o}}bjartsson, A.},
  \bibinfo{author}{Lalu, M.M.}, \bibinfo{author}{Li, T.},
  \bibinfo{author}{Loder, E.W.}, \bibinfo{author}{Mayo-Wilson, E.},
  \bibinfo{author}{McDonald, S.}, \bibinfo{author}{McGuinness, L.A.},
  \bibinfo{author}{Stewart, L.A.}, \bibinfo{author}{Thomas, J.},
  \bibinfo{author}{Tricco, A.C.}, \bibinfo{author}{Welch, V.A.},
  \bibinfo{author}{Whiting, P.}, \bibinfo{author}{Moher, D.},
  \bibinfo{year}{2021}.
\newblock \bibinfo{title}{{The PRISMA 2020 statement: an updated guideline for
  reporting systematic reviews}}.
\newblock \bibinfo{journal}{Systematic Reviews} \bibinfo{volume}{10},
  \bibinfo{pages}{1--11}.
\newblock \DOIprefix\doi{10.1186/s13643-021-01626-4}.
\bibitem[{Palanker(2016)}]{Palanker}
\bibinfo{author}{Palanker}, \bibinfo{year}{2016}.
\newblock \bibinfo{title}{{File:Diagram of the eye and placement of the retinal
  implants.jpg - Wikimedia Commons}}.
\newblock \URLprefix
  \url{https://commons.wikimedia.org/wiki/\\File:Diagram\_of\_the\_eye\_and\_placement\_of\_the\_retinal\_implants.jpg}.
  \bibinfo{note}{[Online; accessed 2022-08-13]}.
\bibitem[{Pan and Chen(2021)}]{Pan2021}
\bibinfo{author}{Pan, L.}, \bibinfo{author}{Chen, X.}, \bibinfo{year}{2021}.
\newblock \bibinfo{title}{{Retinal OCT Image Registration: Methods and
  Applications}}.
\newblock \bibinfo{journal}{IEEE Reviews in Biomedical Engineering}
  \DOIprefix\doi{10.1109/RBME.2021.3110958}.
\bibitem[{Pumpkinegam(2006)}]{Pumpkinegam}
\bibinfo{author}{Pumpkinegam}, \bibinfo{year}{2006}.
\newblock \bibinfo{title}{{File:Full-field OCT setup.png - Wikimedia Commons}}.
\newblock \URLprefix
  \url{https://commons.wikimedia.org/wiki/\\File:Full-field\_OCT\_setup.png}.
  \bibinfo{note}{[Online; accessed 2022-08-13]}.
\bibitem[{Querques et~al.(2019)Querques, Borrelli, Sacconi, {De Vitis},
  Leocani, Santangelo, Magnani, Comi and Bandello}]{Querques2019}
\bibinfo{author}{Querques, G.}, \bibinfo{author}{Borrelli, E.},
  \bibinfo{author}{Sacconi, R.}, \bibinfo{author}{{De Vitis}, L.},
  \bibinfo{author}{Leocani, L.}, \bibinfo{author}{Santangelo, R.},
  \bibinfo{author}{Magnani, G.}, \bibinfo{author}{Comi, G.},
  \bibinfo{author}{Bandello, F.}, \bibinfo{year}{2019}.
\newblock \bibinfo{title}{{Functional and morphological changes of the retinal
  vessels in Alzheimer's disease and mild cognitive impairment}}.
\newblock \bibinfo{journal}{Scientific Reports} \bibinfo{volume}{9}.
\newblock \DOIprefix\doi{10.1038/S41598-018-37271-6}.
\bibitem[{Ran and Cheung(2021)}]{Ran2021a}
\bibinfo{author}{Ran, A.}, \bibinfo{author}{Cheung, C.Y.},
  \bibinfo{year}{2021}.
\newblock \bibinfo{title}{{Deep Learning-Based Optical Coherence Tomography and
  Optical Coherence Tomography Angiography Image Analysis: An Updated
  Summary}}.
\newblock \bibinfo{journal}{Asia-Pacific journal of ophthalmology
  (Philadelphia, Pa.)} \bibinfo{volume}{10}, \bibinfo{pages}{253--260}.
\newblock \DOIprefix\doi{10.1097/APO.0000000000000405}.
\bibitem[{Ran et~al.(2021)Ran, Tham, Chan, Cheng, Tham, Rim and
  Cheung}]{Ran2021}
\bibinfo{author}{Ran, A.R.}, \bibinfo{author}{Tham, C.C.},
  \bibinfo{author}{Chan, P.P.}, \bibinfo{author}{Cheng, C.Y.},
  \bibinfo{author}{Tham, Y.C.}, \bibinfo{author}{Rim, T.H.},
  \bibinfo{author}{Cheung, C.Y.}, \bibinfo{year}{2021}.
\newblock \bibinfo{title}{{Deep learning in glaucoma with optical coherence
  tomography: a review}}.
\newblock \bibinfo{journal}{Eye (London)} \bibinfo{volume}{35},
  \bibinfo{pages}{188--201}.
\newblock \DOIprefix\doi{10.1038/s41433-020-01191-5}.
\bibitem[{Ribeiro et~al.(2016)Ribeiro, Singh and Guestrin}]{Ribeiro2016}
\bibinfo{author}{Ribeiro, M.T.}, \bibinfo{author}{Singh, S.},
  \bibinfo{author}{Guestrin, C.}, \bibinfo{year}{2016}.
\newblock \bibinfo{title}{{"Why Should I Trust You?" Explaining the Predictions
  of Any Alassifier}}.
\newblock \bibinfo{journal}{Proceedings of the ACM SIGKDD International
  Conference on Knowledge Discovery and Data Mining}
  \bibinfo{volume}{13-17-Augu}, \bibinfo{pages}{1135--1144}.
\newblock \DOIprefix\doi{10.1145/2939672.2939778},
  \href{http://arxiv.org/abs/arXiv:1602.04938v3}{{\tt
  arXiv:arXiv:1602.04938v3}}.
\bibitem[{Romaus-Sanjurjo et~al.(2022)Romaus-Sanjurjo, Regueiro,
  L{\'{o}}pez-L{\'{o}}pez, V{\'{a}}zquez-V{\'{a}}zquez, Ouro, Lema and
  Sobrino}]{Romaus-Sanjurjo2022}
\bibinfo{author}{Romaus-Sanjurjo, D.}, \bibinfo{author}{Regueiro, U.},
  \bibinfo{author}{L{\'{o}}pez-L{\'{o}}pez, M.},
  \bibinfo{author}{V{\'{a}}zquez-V{\'{a}}zquez, L.}, \bibinfo{author}{Ouro,
  A.}, \bibinfo{author}{Lema, I.}, \bibinfo{author}{Sobrino, T.},
  \bibinfo{year}{2022}.
\newblock \bibinfo{title}{{Alzheimer's Disease Seen through the Eye: Ocular
  Alterations and Neurodegeneration}}.
\newblock \bibinfo{journal}{International Journal of Molecular Sciences}
  \bibinfo{volume}{23}.
\newblock \DOIprefix\doi{10.3390/ijms23052486}.
\bibitem[{Salobrar-Garc{\'{i}}a et~al.(2019)Salobrar-Garc{\'{i}}a, {De Hoz},
  Ram{\'{i}}rez, L{\'{o}}pez-Cuenca, Rojas, Vazirani, Amarante, Yubero, Gil,
  Pinazo-Dur{\'{a}}n, Salazar and Ram{\'{i}}rez}]{Salobrar-Garcia2019}
\bibinfo{author}{Salobrar-Garc{\'{i}}a, E.}, \bibinfo{author}{{De Hoz}, R.},
  \bibinfo{author}{Ram{\'{i}}rez, A.I.}, \bibinfo{author}{L{\'{o}}pez-Cuenca,
  I.}, \bibinfo{author}{Rojas, P.}, \bibinfo{author}{Vazirani, R.},
  \bibinfo{author}{Amarante, C.}, \bibinfo{author}{Yubero, R.},
  \bibinfo{author}{Gil, P.}, \bibinfo{author}{Pinazo-Dur{\'{a}}n, M.D.},
  \bibinfo{author}{Salazar, J.J.}, \bibinfo{author}{Ram{\'{i}}rez, J.M.},
  \bibinfo{year}{2019}.
\newblock \bibinfo{title}{{Changes in visual function and retinal structure in
  the progression of Alzheimer's disease}}.
\newblock \bibinfo{journal}{PLoS ONE} \bibinfo{volume}{14}.
\newblock \DOIprefix\doi{10.1371/JOURNAL.PONE.0220535}.
\bibitem[{Sampson et~al.(2022)Sampson, Dubis, Chen, Zawadzki and
  Sampson}]{Sampson2022}
\bibinfo{author}{Sampson, D.M.}, \bibinfo{author}{Dubis, A.M.},
  \bibinfo{author}{Chen, F.K.}, \bibinfo{author}{Zawadzki, R.J.},
  \bibinfo{author}{Sampson, D.D.}, \bibinfo{year}{2022}.
\newblock \bibinfo{title}{{Towards standardizing retinal optical coherence
  tomography angiography: a review}}.
\newblock \bibinfo{journal}{Light: Science and Applications}
  \bibinfo{volume}{11}.
\newblock \DOIprefix\doi{10.1038/s41377-022-00740-9}.
\bibitem[{Sandeep et~al.(2017)Sandeep, {Kumar A}, Mahadevan and P}]{CS2017}
\bibinfo{author}{Sandeep, C.S.}, \bibinfo{author}{{Kumar A}, S.},
  \bibinfo{author}{Mahadevan, K.}, \bibinfo{author}{P, M.},
  \bibinfo{year}{2017}.
\newblock \bibinfo{title}{{Dimensionality Reduction of Optical Coherence
  Tomography Images for the Early Diagnosis of Alzheimer's Disease}}.
\newblock \bibinfo{journal}{American Journal of Electrical and Electronic
  Engineering} \bibinfo{volume}{5}, \bibinfo{pages}{58--63}.
\newblock \DOIprefix\doi{10.12691/ajeee-5-2-4}.
\bibitem[{Sandeep et~al.(2019)Sandeep, {Sukesh Kumar}, Mahadevan and
  Manoj}]{Sandeep2019}
\bibinfo{author}{Sandeep, C.S.}, \bibinfo{author}{{Sukesh Kumar}, A.},
  \bibinfo{author}{Mahadevan, K.}, \bibinfo{author}{Manoj, P.},
  \bibinfo{year}{2019}.
\newblock \bibinfo{title}{{Analysis of retinal OCT images for the early
  diagnosis of Alzheimer's disease}}.
\newblock \bibinfo{journal}{Advances in Intelligent Systems and Computing}
  \bibinfo{volume}{749}, \bibinfo{pages}{509--520}.
\newblock \DOIprefix\doi{10.1007/978-3-319-74808-5\_43}.
\bibitem[{Sayeed et~al.(2022)Sayeed, Ahmed, Vinmathi, Priyadarsini, Gundupalli,
  Tripathi, Shishah and Sundramurthy}]{Sayeed2022}
\bibinfo{author}{Sayeed, F.}, \bibinfo{author}{Ahmed, K.R.},
  \bibinfo{author}{Vinmathi, M.S.}, \bibinfo{author}{Priyadarsini, A.I.},
  \bibinfo{author}{Gundupalli, C.B.}, \bibinfo{author}{Tripathi, V.},
  \bibinfo{author}{Shishah, W.}, \bibinfo{author}{Sundramurthy, V.P.},
  \bibinfo{year}{2022}.
\newblock \bibinfo{title}{{Classification of Transgenic Mice by Retinal Imaging
  Using SVMS}}.
\newblock \bibinfo{journal}{Computational Intelligence and Neuroscience}
  \bibinfo{volume}{2022}.
\newblock \DOIprefix\doi{10.1155/2022/9063880}.
\bibitem[{Schott(2020)}]{Schott2020}
\bibinfo{author}{Schott, R.}, \bibinfo{year}{2020}.
\newblock \bibinfo{title}{{How Do OCT Devices for Glaucoma Compare?}}
\newblock \bibinfo{journal}{Review of Optometry} \URLprefix
  \url{https://www.reviewofoptometry.com/article/\\how-do-oct-devices-for-glaucoma-compare}.
\bibitem[{Selvaraju et~al.(2020)Selvaraju, Cogswell, Das, Vedantam, Parikh and
  Batra}]{Selvaraju2020}
\bibinfo{author}{Selvaraju, R.R.}, \bibinfo{author}{Cogswell, M.},
  \bibinfo{author}{Das, A.}, \bibinfo{author}{Vedantam, R.},
  \bibinfo{author}{Parikh, D.}, \bibinfo{author}{Batra, D.},
  \bibinfo{year}{2020}.
\newblock \bibinfo{title}{{Grad-CAM: Visual Explanations from Deep Networks via
  Gradient-Based Localization}}.
\newblock \bibinfo{journal}{International Journal of Computer Vision}
  \bibinfo{volume}{128}, \bibinfo{pages}{336--359}.
\newblock \DOIprefix\doi{10.1007/s11263-019-01228-7},
  \href{http://arxiv.org/abs/1610.02391}{{\tt arXiv:1610.02391}}.
\bibitem[{Singh et~al.(2021)Singh, Balaji, Rasheed, Jayakumar, Raman and
  Lakshminarayanan}]{Singh2021}
\bibinfo{author}{Singh, A.}, \bibinfo{author}{Balaji, J.J.},
  \bibinfo{author}{Rasheed, M.A.}, \bibinfo{author}{Jayakumar, V.},
  \bibinfo{author}{Raman, R.}, \bibinfo{author}{Lakshminarayanan, V.},
  \bibinfo{year}{2021}.
\newblock \bibinfo{title}{{Evaluation of explainable deep learning methods for
  ophthalmic diagnosis}}.
\newblock \bibinfo{journal}{Clinical Ophthalmology} \bibinfo{volume}{15},
  \bibinfo{pages}{2573--2581}.
\newblock \DOIprefix\doi{10.2147/OPTH.S312236}.
\bibitem[{Snyder et~al.(2020)Snyder, Alber, Alt, Bain, Bouma, Bouwman, {Cabrera
  Debuc}, Campbell, Carrillo, Chew, Cordeiro, Due{\~{n}}as, Fern{\'{a}}ndez,
  Koronyo-Hamaoui and Snyder}]{Snyder2020}
\bibinfo{author}{Snyder, P.J.}, \bibinfo{author}{Alber, J.},
  \bibinfo{author}{Alt, C.}, \bibinfo{author}{Bain, L.J.},
  \bibinfo{author}{Bouma, B.E.}, \bibinfo{author}{Bouwman, F.H.},
  \bibinfo{author}{{Cabrera Debuc}, D.}, \bibinfo{author}{Campbell, M.C.W.},
  \bibinfo{author}{Carrillo, M.C.}, \bibinfo{author}{Chew, E.Y.},
  \bibinfo{author}{Cordeiro, M.F.}, \bibinfo{author}{Due{\~{n}}as, M.R.},
  \bibinfo{author}{Fern{\'{a}}ndez, B.M.}, \bibinfo{author}{Koronyo-Hamaoui,
  M.}, \bibinfo{author}{Snyder, H.M.}, \bibinfo{year}{2020}.
\newblock \bibinfo{title}{{Retinal imaging in Alzheimer's and neurodegenerative
  diseases}}.
\newblock \bibinfo{journal}{Alzheimer's \& Dementia}
  \DOIprefix\doi{10.1002/alz.12179}.
\bibitem[{Song et~al.(2021)Song, Johnson, Ayala and Thompson}]{Song2021}
\bibinfo{author}{Song, A.}, \bibinfo{author}{Johnson, N.},
  \bibinfo{author}{Ayala, A.}, \bibinfo{author}{Thompson, A.C.},
  \bibinfo{year}{2021}.
\newblock \bibinfo{title}{{Optical coherence tomography in patients with
  alzheimer's disease: What can it tell us?}}
\newblock \bibinfo{journal}{Eye and Brain} \bibinfo{volume}{13},
  \bibinfo{pages}{1--20}.
\newblock \DOIprefix\doi{10.2147/EB.S235238}.
\bibitem[{Thakoor et~al.(2022)Thakoor, Yao, Bordbar, Moussa, Lin, Sajda and
  Chen}]{Thakoor2022}
\bibinfo{author}{Thakoor, K.A.}, \bibinfo{author}{Yao, J.},
  \bibinfo{author}{Bordbar, D.}, \bibinfo{author}{Moussa, O.},
  \bibinfo{author}{Lin, W.}, \bibinfo{author}{Sajda, P.},
  \bibinfo{author}{Chen, R.W.}, \bibinfo{year}{2022}.
\newblock \bibinfo{title}{{A multimodal deep learning system to distinguish
  late stages of AMD and to compare expert vs. AI ocular biomarkers}}.
\newblock \bibinfo{journal}{Scientific Reports} \bibinfo{volume}{12},
  \bibinfo{pages}{1--11}.
\newblock \URLprefix \url{https://doi.org/10.1038/s41598-022-06273-w},
  \DOIprefix\doi{10.1038/s41598-022-06273-w}.
\bibitem[{Tian et~al.(2021a)Tian, Smith, Guo, Liu, Pan, Wang, Xiong and
  Fang}]{Tian2021}
\bibinfo{author}{Tian, J.}, \bibinfo{author}{Smith, G.}, \bibinfo{author}{Guo,
  H.}, \bibinfo{author}{Liu, B.}, \bibinfo{author}{Pan, Z.},
  \bibinfo{author}{Wang, Z.}, \bibinfo{author}{Xiong, S.},
  \bibinfo{author}{Fang, R.}, \bibinfo{year}{2021}a.
\newblock \bibinfo{title}{{Modular machine learning for Alzheimer's disease
  classification from retinal vasculature}}.
\newblock \bibinfo{journal}{Scientific Reports |} \bibinfo{volume}{11},
  \bibinfo{pages}{238}.
\newblock \URLprefix \url{https://doi.org/10.1038/s41598-020-80312-2},
  \DOIprefix\doi{10.1038/s41598-020-80312-2}.
\bibitem[{Tian et~al.(2021b)Tian, Hunt, Bell, Yi, Smith, Ochoa, Intes and
  Durr}]{Tian2021a}
\bibinfo{author}{Tian, L.}, \bibinfo{author}{Hunt, B.}, \bibinfo{author}{Bell,
  M.A.}, \bibinfo{author}{Yi, J.}, \bibinfo{author}{Smith, J.T.},
  \bibinfo{author}{Ochoa, M.}, \bibinfo{author}{Intes, X.},
  \bibinfo{author}{Durr, N.J.}, \bibinfo{year}{2021}b.
\newblock \bibinfo{title}{{Deep Learning in Biomedical Optics}}.
\newblock \bibinfo{journal}{Lasers in Surgery and Medicine}
  \bibinfo{volume}{53}, \bibinfo{pages}{748--775}.
\newblock \DOIprefix\doi{10.1002/LSM.23414},
  \href{http://arxiv.org/abs/2105.11046}{{\tt arXiv:2105.11046}}.
\bibitem[{Tong et~al.(2020)Tong, Lu, Yu and Shen}]{Tong2020}
\bibinfo{author}{Tong, Y.}, \bibinfo{author}{Lu, W.}, \bibinfo{author}{Yu, Y.},
  \bibinfo{author}{Shen, Y.}, \bibinfo{year}{2020}.
\newblock \bibinfo{title}{{Application of machine learning in ophthalmic
  imaging modalities}}.
\newblock \bibinfo{journal}{Eye and vision (Novato, Calif.)}
  \bibinfo{volume}{7}.
\newblock \DOIprefix\doi{10.1186/s40662-020-00183-6}.
\bibitem[{Trebbastoni et~al.(2017)Trebbastoni, Marcelli, Mallone, D'Antonio,
  Imbriano, Campanelli, {De Lena} and Gharbiya}]{Trebbastoni2017}
\bibinfo{author}{Trebbastoni, A.}, \bibinfo{author}{Marcelli, M.},
  \bibinfo{author}{Mallone, F.}, \bibinfo{author}{D'Antonio, F.},
  \bibinfo{author}{Imbriano, L.}, \bibinfo{author}{Campanelli, A.},
  \bibinfo{author}{{De Lena}, C.}, \bibinfo{author}{Gharbiya, M.},
  \bibinfo{year}{2017}.
\newblock \bibinfo{title}{{Attenuation of Choroidal Thickness in Patients With
  Alzheimer Disease: Evidence From an Italian Prospective Study}}.
\newblock \bibinfo{journal}{Alzheimer Disease and Associated Disorders}
  \bibinfo{volume}{31}, \bibinfo{pages}{128--134}.
\newblock \DOIprefix\doi{10.1097/WAD.0000000000000176}.
\bibitem[{Trindade(2021)}]{Trindade}
\bibinfo{author}{Trindade, R.C.}, \bibinfo{year}{2021}.
\newblock \bibinfo{title}{{Identification of Changes in Retinal Images of
  Animal Models of Alzheimer's Disease Using Deep Learning Approach}}.
\newblock Master's thesis. Iniversity of Coimbra.
\bibitem[{Triolo et~al.(2017)Triolo, Rabiolo, Shemonski, Fard, {Di Matteo},
  Sacconi, Bettin, Magazzeni, Querques, Vazquez, Barboni and
  Bandello}]{Triolo2017}
\bibinfo{author}{Triolo, G.}, \bibinfo{author}{Rabiolo, A.},
  \bibinfo{author}{Shemonski, N.D.}, \bibinfo{author}{Fard, A.},
  \bibinfo{author}{{Di Matteo}, F.}, \bibinfo{author}{Sacconi, R.},
  \bibinfo{author}{Bettin, P.}, \bibinfo{author}{Magazzeni, S.},
  \bibinfo{author}{Querques, G.}, \bibinfo{author}{Vazquez, L.E.},
  \bibinfo{author}{Barboni, P.}, \bibinfo{author}{Bandello, F.},
  \bibinfo{year}{2017}.
\newblock \bibinfo{title}{{Optical coherence tomography angiography macular and
  peripapillary vessel perfusion density in healthy subjects, glaucoma
  suspects, and glaucoma patients}}.
\newblock \bibinfo{journal}{Investigative Ophthalmology and Visual Science}
  \bibinfo{volume}{58}, \bibinfo{pages}{5713--5722}.
\newblock \DOIprefix\doi{10.1167/iovs.17-22865}.
\bibitem[{Tsang and Sharma(2018)}]{Tsang2018}
\bibinfo{author}{Tsang, S.H.}, \bibinfo{author}{Sharma, T.},
  \bibinfo{year}{2018}.
\newblock \bibinfo{title}{{Electroretinography}}.
\newblock \bibinfo{publisher}{Springer International Publishing},
  \bibinfo{address}{Cham}.
\newblock \URLprefix \url{https://doi.org/10.1007/978-3-319-95046-4\_5},
  \DOIprefix\doi{10.1007/978-3-319-95046-4\_5}.
\bibitem[{{UK Biobank}(2022)}]{UK_Biobank}
\bibinfo{author}{{UK Biobank}}, \bibinfo{year}{2022}.
\newblock \bibinfo{title}{{UK Biobank - UK Biobank}}.
\newblock \URLprefix \url{https://www.ukbiobank.ac.uk/}.
  \bibinfo{note}{[Online; accessed 2022-08-13]}.
\bibitem[{Vij and Arora(2022)}]{Vij2022}
\bibinfo{author}{Vij, R.}, \bibinfo{author}{Arora, S.}, \bibinfo{year}{2022}.
\newblock \bibinfo{title}{{A systematic survey of advances in retinal imaging
  modalities for Alzheimer's disease diagnosis}}.
\newblock \bibinfo{number}{0123456789}, \bibinfo{publisher}{Springer US}.
\newblock \URLprefix \url{https://doi.org/10.1007/s11011-022-00927-4},
  \DOIprefix\doi{10.1007/s11011-022-00927-4}.
\bibitem[{Wada et~al.(2020)Wada, Song, Oomae, Sogawa, Yoshioka, Nakabayashi,
  Takahashi, Tani, Ishibazawa, Ishiko and Yoshida}]{Wada2020}
\bibinfo{author}{Wada, T.}, \bibinfo{author}{Song, Y.}, \bibinfo{author}{Oomae,
  T.}, \bibinfo{author}{Sogawa, K.}, \bibinfo{author}{Yoshioka, T.},
  \bibinfo{author}{Nakabayashi, S.}, \bibinfo{author}{Takahashi, K.},
  \bibinfo{author}{Tani, T.}, \bibinfo{author}{Ishibazawa, A.},
  \bibinfo{author}{Ishiko, S.}, \bibinfo{author}{Yoshida, A.},
  \bibinfo{year}{2020}.
\newblock \bibinfo{title}{{Longitudinal changes in retinal blood flow in a
  feline retinal vein occlusion model as measured by Doppler optical coherence
  tomography and optical coherence tomography angiography}}.
\newblock \bibinfo{journal}{Investigative Ophthalmology and Visual Science}
  \bibinfo{volume}{61}.
\newblock \DOIprefix\doi{10.1167/iovs.61.2.34}.
\bibitem[{Wagner et~al.(2020)Wagner, Fu, Faes, Liu, Huemer, Khalid, Ferraz,
  Korot, Kelly, Balaskas, Denniston and Keane}]{Wagner2020}
\bibinfo{author}{Wagner, S.K.}, \bibinfo{author}{Fu, D.J.},
  \bibinfo{author}{Faes, L.}, \bibinfo{author}{Liu, X.},
  \bibinfo{author}{Huemer, J.}, \bibinfo{author}{Khalid, H.},
  \bibinfo{author}{Ferraz, D.}, \bibinfo{author}{Korot, E.},
  \bibinfo{author}{Kelly, C.}, \bibinfo{author}{Balaskas, K.},
  \bibinfo{author}{Denniston, A.K.}, \bibinfo{author}{Keane, P.A.},
  \bibinfo{year}{2020}.
\newblock \bibinfo{title}{{Insights into systemic disease through retinal
  imaging-based oculomics}}.
\newblock \bibinfo{journal}{Translational Vision Science and Technology}
  \bibinfo{volume}{9}.
\newblock \DOIprefix\doi{10.1167/tvst.9.2.6}.
\bibitem[{Wagner et~al.(2022)Wagner, Hughes, Cortina-Borja, Pontikos, Struyven,
  Liu, Montgomery, Alexander, Topol, Petersen, Balaskas, Hindley, Petzold,
  Rahi, Denniston and Keane}]{Wagner2022}
\bibinfo{author}{Wagner, S.K.}, \bibinfo{author}{Hughes, F.},
  \bibinfo{author}{Cortina-Borja, M.}, \bibinfo{author}{Pontikos, N.},
  \bibinfo{author}{Struyven, R.}, \bibinfo{author}{Liu, X.},
  \bibinfo{author}{Montgomery, H.}, \bibinfo{author}{Alexander, D.C.},
  \bibinfo{author}{Topol, E.}, \bibinfo{author}{Petersen, S.E.},
  \bibinfo{author}{Balaskas, K.}, \bibinfo{author}{Hindley, J.},
  \bibinfo{author}{Petzold, A.}, \bibinfo{author}{Rahi, J.S.},
  \bibinfo{author}{Denniston, A.K.}, \bibinfo{author}{Keane, P.A.},
  \bibinfo{year}{2022}.
\newblock \bibinfo{title}{{AlzEye: longitudinal record-level linkage of
  ophthalmic imaging and hospital admissions of 353157 patients in London,
  UK}}.
\newblock \bibinfo{journal}{BMJ open} \bibinfo{volume}{12},
  \bibinfo{pages}{e058552}.
\newblock \DOIprefix\doi{10.1136/bmjopen-2021-058552}.
\bibitem[{{WHO}(2014)}]{WorldHealthOrganization}
\bibinfo{author}{{WHO}}, \bibinfo{year}{2014}.
\newblock \bibinfo{title}{{International statistical classification of diseases
  and related health problems 10th Revision.}}
\newblock \bibinfo{publisher}{World Health Organization}.
\newblock \URLprefix \url{https://icd.who.int/browse10/2014/en\#/F00\-F09}.
\bibitem[{WHO(2019)}]{WHO2019}
\bibinfo{author}{WHO}, \bibinfo{year}{2019}.
\newblock \bibinfo{title}{{World Health Organization: Dementia Fact Sheet}}.
\newblock \URLprefix
  \url{https://www.who.int/news-room/fact-sheets/detail/dementia}.
\bibitem[{Wisely et~al.(2022)Wisely, Wang, Henao, Grewal, Thompson, Robbins,
  Yoon, Soundararajan, Polascik, Burke, Liu, Carin and Fekrat}]{Wisely2022}
\bibinfo{author}{Wisely, C.E.}, \bibinfo{author}{Wang, D.},
  \bibinfo{author}{Henao, R.}, \bibinfo{author}{Grewal, D.S.},
  \bibinfo{author}{Thompson, A.C.}, \bibinfo{author}{Robbins, C.B.},
  \bibinfo{author}{Yoon, S.P.}, \bibinfo{author}{Soundararajan, S.},
  \bibinfo{author}{Polascik, B.W.}, \bibinfo{author}{Burke, J.R.},
  \bibinfo{author}{Liu, A.}, \bibinfo{author}{Carin, L.},
  \bibinfo{author}{Fekrat, S.}, \bibinfo{year}{2022}.
\newblock \bibinfo{title}{{Convolutional neural network to identify symptomatic
  Alzheimer's disease using multimodal retinal imaging}}.
\newblock \bibinfo{journal}{British Journal of Ophthalmology}
  \bibinfo{volume}{106}, \bibinfo{pages}{388--395}.
\newblock \DOIprefix\doi{10.1136/bjophthalmol-2020-317659}.
\bibitem[{WorldMedicalAssociation(2013)}]{helsinki2013}
\bibinfo{author}{WorldMedicalAssociation}, \bibinfo{year}{2013}.
\newblock \bibinfo{title}{{World Medical Association declaration of Helsinki:
  Ethical principles for medical research involving human subjects}}.
\newblock \DOIprefix\doi{10.1001/jama.2013.281053}.
\bibitem[{Yamanakkanavar et~al.(2020)Yamanakkanavar, Choi and
  Lee}]{Yamanakkanavar2020}
\bibinfo{author}{Yamanakkanavar, N.}, \bibinfo{author}{Choi, J.Y.},
  \bibinfo{author}{Lee, B.}, \bibinfo{year}{2020}.
\newblock \bibinfo{title}{{MRI segmentation and classification of human brain
  using deep learning for diagnosis of alzheimer's disease: A survey}}.
\newblock \DOIprefix\doi{10.3390/s20113243}.
\bibitem[{Yan et~al.(2021)Yan, Jin, Gao, Huang, Wang, Wang and Ye}]{Yan2021}
\bibinfo{author}{Yan, Y.}, \bibinfo{author}{Jin, K.}, \bibinfo{author}{Gao,
  Z.}, \bibinfo{author}{Huang, X.}, \bibinfo{author}{Wang, F.},
  \bibinfo{author}{Wang, Y.}, \bibinfo{author}{Ye, J.}, \bibinfo{year}{2021}.
\newblock \bibinfo{title}{{Attention-based deep learning system for automated
  diagnoses of age-related macular degeneration in optical coherence tomography
  images}}.
\newblock \bibinfo{journal}{Medical Physics} \bibinfo{volume}{48},
  \bibinfo{pages}{4926--4934}.
\newblock \DOIprefix\doi{10.1002/MP.15002}.
\bibitem[{Yanagihara et~al.(2020)Yanagihara, Lee, Ting and
  Lee}]{Yanagihara2020}
\bibinfo{author}{Yanagihara, R.T.}, \bibinfo{author}{Lee, C.S.},
  \bibinfo{author}{Ting, D.S.W.}, \bibinfo{author}{Lee, A.Y.},
  \bibinfo{year}{2020}.
\newblock \bibinfo{title}{{Methodological challenges of deep learning in
  optical coherence tomography for retinal diseases: A review}}.
\newblock \bibinfo{journal}{Translational Vision Science and Technology}
  \bibinfo{volume}{9}, \bibinfo{pages}{17--19}.
\newblock \DOIprefix\doi{10.1167/tvst.9.2.11}.
\bibitem[{Yeh et~al.(2022)Yeh, Kuo and Chou}]{Yeh2022}
\bibinfo{author}{Yeh, T.C.}, \bibinfo{author}{Kuo, C.T.},
  \bibinfo{author}{Chou, Y.B.}, \bibinfo{year}{2022}.
\newblock \bibinfo{title}{{Retinal Microvascular Changes in Mild Cognitive
  Impairment and Alzheimer's Disease: A Systematic Review, Meta-Analysis, and
  Meta-Regression}}.
\newblock \bibinfo{journal}{Frontiers in Aging Neuroscience}
  \bibinfo{volume}{14}.
\newblock \DOIprefix\doi{10.3389/fnagi.2022.860759}.
\bibitem[{Yoon et~al.(2019)Yoon, Thompson, Polascik, Calixte, Burke, Petrella,
  Grewal and Fekrat}]{Yoon2019}
\bibinfo{author}{Yoon, S.P.}, \bibinfo{author}{Thompson, A.C.},
  \bibinfo{author}{Polascik, B.W.}, \bibinfo{author}{Calixte, C.},
  \bibinfo{author}{Burke, J.R.}, \bibinfo{author}{Petrella, J.R.},
  \bibinfo{author}{Grewal, D.S.}, \bibinfo{author}{Fekrat, S.},
  \bibinfo{year}{2019}.
\newblock \bibinfo{title}{{Correlation of OctA and volumetric MRI in mild
  cognitive impairment and Alzheimer's disease}}.
\newblock \bibinfo{journal}{Ophthalmic Surgery Lasers and Imaging Retina}
  \bibinfo{volume}{50}, \bibinfo{pages}{709--718}.
\newblock \DOIprefix\doi{10.3928/23258160-20191031-06}.
\bibitem[{Yuan et~al.(2022)Yuan, Lee and Ms}]{Yuan2022}
\bibinfo{author}{Yuan, A.}, \bibinfo{author}{Lee, C.S.}, \bibinfo{author}{Ms,
  {\~{A}}.}, \bibinfo{year}{2022}.
\newblock \bibinfo{title}{{Retinal Biomarkers for Alzheimer Disease : The Facts
  and the Future}}.
\newblock \bibinfo{journal}{Asia-Pasific Journal of Ophthalmology} ,
  \bibinfo{pages}{140--148}\DOIprefix\doi{10.1097/APO.0000000000000505}.
\bibitem[{Zeng et~al.(2021)Zeng, Han, Zhang and Bai}]{Zeng2021}
\bibinfo{author}{Zeng, H.M.}, \bibinfo{author}{Han, H.B.},
  \bibinfo{author}{Zhang, Q.F.}, \bibinfo{author}{Bai, H.},
  \bibinfo{year}{2021}.
\newblock \bibinfo{title}{{Application of modern neuroimaging technology in the
  diagnosis and study of Alzheimer's disease}}.
\newblock \DOIprefix\doi{10.4103/1673-5374.286957}.
\bibitem[{Zhang et~al.(2022)Zhang, Gong, Ma, Wen, Wang and Yao}]{Zhang2022}
\bibinfo{author}{Zhang, M.}, \bibinfo{author}{Gong, X.}, \bibinfo{author}{Ma,
  W.}, \bibinfo{author}{Wen, L.}, \bibinfo{author}{Wang, Y.},
  \bibinfo{author}{Yao, H.}, \bibinfo{year}{2022}.
\newblock \bibinfo{title}{{A Study on the Correlation Between Age-Related
  Macular Degeneration and Alzheimer ' s Disease Based on the Application of
  Artificial Neural Network}}.
\newblock \bibinfo{journal}{Frontiers in Public Health} \bibinfo{volume}{10},
  \bibinfo{pages}{1--12}.
\newblock \DOIprefix\doi{10.3389/fpubh.2022.925147}.
\bibitem[{Zhang et~al.(2021)Zhang, Li, Bian, He, Shen, Lan and
  Huang}]{Zhang2021}
\bibinfo{author}{Zhang, Q.}, \bibinfo{author}{Li, J.}, \bibinfo{author}{Bian,
  M.}, \bibinfo{author}{He, Q.}, \bibinfo{author}{Shen, Y.},
  \bibinfo{author}{Lan, Y.}, \bibinfo{author}{Huang, D.}, \bibinfo{year}{2021}.
\newblock \bibinfo{title}{{Retinal imaging techniques based on machine learning
  models in recognition and prediction of mild cognitive impairment}}.
\newblock \bibinfo{journal}{Neuropsychiatric Disease and Treatment}
  \bibinfo{volume}{17}, \bibinfo{pages}{3267--3281}.
\newblock \DOIprefix\doi{10.2147/NDT.S333833}.
\bibitem[{Zhang et~al.(2014)Zhang, Srivastava, Liu, Chen, Duan, {Kee Wong},
  Kwoh, Wong and Liu}]{Zhang2014}
\bibinfo{author}{Zhang, Z.}, \bibinfo{author}{Srivastava, R.},
  \bibinfo{author}{Liu, H.}, \bibinfo{author}{Chen, X.}, \bibinfo{author}{Duan,
  L.}, \bibinfo{author}{{Kee Wong}, D.W.}, \bibinfo{author}{Kwoh, C.K.},
  \bibinfo{author}{Wong, T.Y.}, \bibinfo{author}{Liu, J.},
  \bibinfo{year}{2014}.
\newblock \bibinfo{title}{{A survey on computer aided diagnosis for ocular
  diseases}}.
\newblock \bibinfo{journal}{BMC Med Inform Decis Mak.} \bibinfo{volume}{14},
  \bibinfo{pages}{80}.
\newblock \DOIprefix\doi{10.1186/1472-6947-14-80}.
\bibitem[{Zhou and Sinai(2006)}]{Et2006}
\bibinfo{author}{Zhou, A.}, \bibinfo{author}{Sinai, M.J.},
  \bibinfo{year}{2006}.
\newblock \bibinfo{title}{Method and system for detecting the effects of
  azheimer's disease in the human retina}.
\newblock \bibinfo{note}{US Patent 6,988,995 B2}.
\bibitem[{Zhou(2010)}]{Zhou2010}
\bibinfo{author}{Zhou, P.}, \bibinfo{year}{2010}.
\newblock \bibinfo{title}{Eye care device and method for assessing and
  predicting dementia}.
\newblock \bibinfo{note}{US Patent 11,288,801 B2}.

\end{thebibliography}





\end{document}